\documentclass[aps,prc,preprint,superscriptaddress,showpacs,showkeys]{revtex4}
\usepackage{amsfonts}
\usepackage[fleqn]{amsmath}  
\usepackage{amssymb}
\usepackage{array}
\usepackage{rotating}
\usepackage{graphicx}
\usepackage{bm}

\begin{document}

\title{Quasifree $\Lambda$, $\Sigma^0$, and $\Sigma^-$ electroproduction
from $^{1,2}$H, $^{3,4}$He, and carbon}

\author{F.~Dohrmann}
\email{F.Dohrmann@fzd.de}
\affiliation{Forschungszentrum Dresden-Rossendorf, 01314 Dresden, Germany}
\affiliation{Argonne National Laboratory, Argonne, Illinois 60439}
\author{A.~Ahmidouch}
\affiliation{Hampton University, Hampton, Virginia 23668}
\affiliation{Kent State University, Kent, Ohio 44242}
\author{C.S.~Armstrong}
\affiliation{Thomas Jefferson National Accelerator Facility, Newport News, Virginia 23606}
\affiliation{College of William and Mary, Williamsburg, Virginia 23187}
\author{J.~Arrington}
\affiliation{Argonne National Laboratory, Argonne, Illinois 60439}
\author{R.~Asaturyan}
\affiliation{Yerevan Physics Institute, Yerevan, Armenia}
\author{S.~Avery}
\affiliation{Hampton University, Hampton, Virginia 23668}
\author{K.~Bailey}
\affiliation{Argonne National Laboratory, Argonne, Illinois 60439}
\author{H.~Bitao}
\affiliation{Hampton University, Hampton, Virginia 23668}
\author{H.~Breuer}
\affiliation{University of Maryland, College Park, Maryland 20742}
\author{D.S.~Brown}
\affiliation{University of Maryland, College Park, Maryland 20742}
\author{R.~Carlini}
\affiliation{Thomas Jefferson National Accelerator Facility, Newport News, Virginia 23606}
\author{J.~Cha}
\affiliation{Hampton University, Hampton, Virginia 23668}
\author{N.~Chant}
\affiliation{University of Maryland, College Park, Maryland 20742}
\author{E.~Christy}
\affiliation{Hampton University, Hampton, Virginia 23668}
\author{A.~Cochran}
\affiliation{Hampton University, Hampton, Virginia 23668}
\author{L.~Cole}
\affiliation{Hampton University, Hampton, Virginia 23668}
\author{J.~Crowder}
\affiliation{Juniata College, Huntingdon, Pennsylvania 16652}
\author{S.~Danagoulian}
\affiliation{North Carolina A\&T State University, Greensboro, North Carolina 27411}
\affiliation{Thomas Jefferson National Accelerator Facility, Newport News, Virginia 23606}
\author{M.~Elaasar}
\affiliation{Southern University at New Orleans, New Orleans, Louisiana 70126}
\author{R.~Ent}
\affiliation{Thomas Jefferson National Accelerator Facility, Newport News, Virginia 23606}
\author{H.~Fenker}
\affiliation{Thomas Jefferson National Accelerator Facility, Newport News, Virginia 23606}
\author{Y.~Fujii}
\affiliation{Tohoku University, Sendai, 980-8577 Japan}
\author{L.~Gan}
\affiliation{Hampton University, Hampton, Virginia 23668}
\author{K.~Garrow}
\affiliation{Thomas Jefferson National Accelerator Facility, Newport News, Virginia 23606}
\author{D.F.~Geesaman}
\affiliation{Argonne National Laboratory, Argonne, Illinois 60439}
\author{P.~Gueye}
\affiliation{Hampton University, Hampton, Virginia 23668}
\author{K.~Hafidi}
\affiliation{Argonne National Laboratory, Argonne, Illinois 60439}
\author{W.~Hinton}
\affiliation{Hampton University, Hampton, Virginia 23668}
\author{H.~Juengst}
\affiliation{University of Minnesota, Minneapolis, Minnesota 55455}
\author{C.~Keppel}
\affiliation{Hampton University, Hampton, Virginia 23668}
\author{Y.~Liang}
\affiliation{Hampton University, Hampton, Virginia 23668}
\author{J.H.~Liu}
\affiliation{University of Minnesota, Minneapolis, Minnesota 55455}
\author{A.~Lung}
\affiliation{Thomas Jefferson National Accelerator Facility, Newport News, Virginia 23606}
\author{D.~Mack}
\affiliation{Thomas Jefferson National Accelerator Facility, Newport News, Virginia 23606}
\author{P.~Markowitz}
\affiliation{Florida International University, Miami, Florida 33199}
\affiliation{Thomas Jefferson National Accelerator Facility, Newport News, Virginia 23606}
\author{J.~Mitchell}
\affiliation{Thomas Jefferson National Accelerator Facility, Newport News, Virginia 23606}
\author{T.~Miyoshi}
\affiliation{Tohoku University, Sendai, 980-8577 Japan}
\author{H.~Mkrtchyan}
\affiliation{Yerevan Physics Institute, Yerevan, Armenia}
\author{S.K.~Mtingwa}
\affiliation{North Carolina A\&T State University, Greensboro, North Carolina 27411}
\author{B.~Mueller}
\affiliation{Argonne National Laboratory, Argonne, Illinois 60439}
\author{G.~Niculescu}
\affiliation{Hampton University, Hampton, Virginia 23668}
\affiliation{Ohio University, Athens, Ohio 45701}
\author{I.~Niculescu}
\affiliation{Hampton University, Hampton, Virginia 23668}
\affiliation{The George Washington University, Washington DC 20052}
\author{D.~Potterveld}
\affiliation{Argonne National Laboratory, Argonne, Illinois 60439}
\author{B.A.~Raue}
\affiliation{Florida International University, Miami, Florida 33199}
\affiliation{Thomas Jefferson National Accelerator Facility, Newport News, Virginia 23606}
\author{P.E.~Reimer}
\affiliation{Argonne National Laboratory, Argonne, Illinois 60439}
\author{J.~Reinhold}
\affiliation{Florida International University, Miami, Florida 33199}
\affiliation{Thomas Jefferson National Accelerator Facility, Newport News, Virginia 23606}
\author{J.~Roche}
\affiliation{College of William and Mary, Williamsburg, Virginia 23187}
\author{M.~Sarsour}
\affiliation{University of Houston, Houston, Texas 77204}
\author{Y.~Sato}
\affiliation{Tohoku University, Sendai, 980-8577 Japan}
\author{R.E.~Segel}
\affiliation{Northwestern University, Evanston, Illinois 60201}
\author{A.~Semenov}
\affiliation{Kent State University, Kent, Ohio 44242}
\author{S.~Stepanyan}
\affiliation{Yerevan Physics Institute, Yerevan, Armenia}
\author{V.~Tadevosyan}
\affiliation{Yerevan Physics Institute, Yerevan, Armenia}
\author{S.~Tajima}
\affiliation{Duke University and Triangle Universities Nuclear Laboratory, Durham, North Carolina 27708}
\author{L.~Tang}
\affiliation{Hampton University, Hampton, Virginia 23668}
\author{A.~Uzzle}
\affiliation{Hampton University, Hampton, Virginia 23668}
\author{S.~Wood}
\affiliation{Thomas Jefferson National Accelerator Facility, Newport News, Virginia 23606}
\author{H.~Yamaguchi}
\affiliation{Tohoku University, Sendai, 980-8577 Japan}
\author{C.~Yan}
\affiliation{Kent State University, Kent, Ohio 44242}
\author{L.~Yuan}
\affiliation{Hampton University, Hampton, Virginia 23668}
\author{B.~Zeidman}
\affiliation{Argonne National Laboratory, Argonne, Illinois 60439}
\author{M.~Zeier}
\affiliation{University of Virginia, Charlottesville, Virginia 22901}
\author{B.~Zihlmann}
\affiliation{University of Virginia, Charlottesville, Virginia 22901}

\date{\today}
\begin{abstract}
  Kaon electroproduction from light nuclei and hydrogen, using
  ${}^1$H, $^2$H, $^3$He, $^4$He, and carbon targets has
  been measured at Jefferson Laboratory. The quasifree angular
  distributions of $\Lambda$ and $\Sigma$ hyperons were determined at
  $Q^2= 0.35$~(GeV/c)$^2$ and $W= 1.91$~GeV.  Electroproduction on
  hydrogen was measured at the same kinematics for reference.
\end{abstract}

\pacs{21.45.+v; 21.80.+a; 25.30.Rw}

\maketitle

\section*{Introduction}
A comprehensive study of kaon electroproduction on light nuclei has
been conducted in Hall C of Thomas Jefferson National Accelerator
Facility (Jefferson Lab or JLab). Data were obtained using electron beams of
3.245 GeV impinging on special high density cryogenic targets for,
$^{1,2}$H, $^{3,4}$He, as well as on a solid carbon target.

Until recently the data base of cross sections of electro- and
photoproduction of strangeness was sparse.  In the case of
photoproduction, considerable amounts of new high quality data for the
proton have been published from experiments at JLab, ELSA, SPring-8,
GRAAL and LNS (cf.~\cite{Dohrmann:2006dx} for a list of references).
These data include cross sections, polarization asymmetries, tensor
polarizations, and decay angle distributions. However, the data base
for photoproduction on nuclei and thus implicitly the neutron remains
scarce (cf.~\cite{Kohri:2006yx, Niculescu:2001rg}).  Only few older
measurements have been reported on
deuterium~\cite{Boyarski:1975dj,Quinn:1979zp} and
carbon~\cite{Yamazaki:1995bc} targets.

Traditionally, $^2$H and $^3$He targets have been considered to be a
good approximation for a free neutron target. In the present work, as
in the majority of kaon electroproduction experiments, a positive kaon
is detected in coincidence with the scattered electron. On the proton,
this leads to two possible final states with either a $\Lambda$ of
$\Sigma^0$ hyperon, that are easily separable by a missing mass
analysis. On the neutron, a $\Sigma^-$ is produced as final state. Due
to the small mass difference of $\Sigma^-$ and $\Sigma^0$ of 4.8
MeV/c$^2$ and the initial nucleon momentum distribution, the $\Sigma$
contributions from the proton and neutron cannot be separated by
missing mass. With increasing target mass, the separation between
$\Lambda$ and $\Sigma$ distributions also gets worse because of the
increasing Fermi momentum. Thus, $^2$H and $^3$He targets offer the
best access to the neutron cross sections. Since a missing mass
analysis, strictly speaking, can only determine the total $\Sigma$
strength, the different $N/Z$ ratio for the $^2$H and $^3$He targets
should assist in further disentangling the $\Sigma^0$ and $\Sigma^-$
contributions.

Systematic studies of heavier nuclei will then provide the
possibilities of investigating in-medium modifications of the
elementary kaon electroproduction mechanism as well as the propagation
of the outgoing $K^+$. e.g.~experimental data on $^{12}$C
\cite{Yamazaki:1995bc,Hinton:1998tr,Miyoshi:2002ts,Yuan:2004dd} show
an effective proton number that is in disagreement with theoretical
calculations~\cite{Lee:1997mt}, thereby indicating the need for
modifications.

We present here the results of an experiment on the electroproduction
of open strangeness on light nuclei with $A=2,3,4,12$, that 
has been performed in Hall C at Jefferson Lab. Also
measured was the production on a hydrogen target. This facilitates
direct comparison to the elementary $p(e,e'K^+)$ reaction for
identical kinematics. Results of this experiment on the production of
$\Lambda$ hypernuclear states, $^3_{\Lambda}\mathrm{H}$ and
$^4_{\Lambda}\mathrm{H}$, have been presented in
Ref.~\cite{Dohrmann:2004xy}.  In this paper we present the cross
sections for the quasifree production of $\Lambda$, $\Sigma^0$,
$\Sigma^-$. To the best knowledge of the authors, this is the first
reported kaon electroproduction measurement on helium isotopes.

\section*{Experiment}
Experiment E91-016 had two runs, one that only used
Hydrogen and Deuterium targets, and a subsequent one that also included
helium and carbon targets.  We present cross sections from the second run, which
included data for all four few-body nuclei. Data were obtained using
electron beams of 3.245 GeV impinging on special high density
cryogenic targets for $^{1,2}$H, $^{3,4}$He. The target thicknesses
were 289 mg/cm$^2$ for $^1$H at 19 K, 668 mg/cm$^2$ for $^2$H at 22 K,
310 mg/cm$^2$ for $^3$He at 5.5 K, and 546 mg/cm$^2$ for $^4$He at 5.5
K.
The target lengths were approximately 4 cm for each target. In
addition, data was taken on a 227 mg/cm$^2$ carbon target.

\begin{figure}[phbt]
  \centering
  \includegraphics[width=12cm]{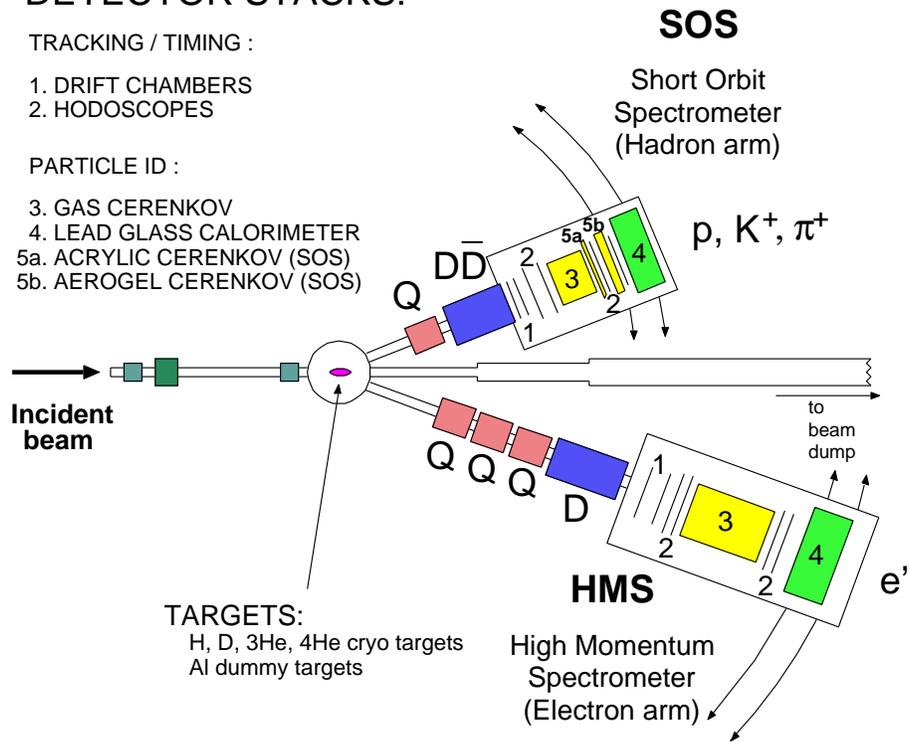}  
  \caption{(Color Online) Setup of the experiment (modified
    from~\cite{Mohring:2002tr,Ambrozewicz:2004kn}). While the general setup was
    similar to other Hall C experiments, in this experiment an
    additional acrylic \v Cerenkov detector was used for better $K^+/p$
    discrimination.}
  \label{fig:hallc}
\end{figure}

The scattered electrons were detected in the High Momentum
Spectrometer (HMS, momentum acceptance $\Delta p/p \approx \pm 10\%$,
solid angle $\approx 6.7$~msr) in coincidence with the electroproduced
kaons, detected in the Short Orbit Spectrometer (SOS, momentum
acceptance $\Delta p/p \approx \pm 20\%$, solid angle $\approx
7.5$~msr).  The detectors and coincidence methods have been described
in detail for similar experiments in Hall C~\cite{Mohring:2002tr,
  Gaskell:2001xr, Dutta:2003yt}.  The detector packages of the two
spectrometers are very similar, and a sketch of the setup of the
experiment is shown in Fig.~\ref{fig:hallc}.  Two drift chambers near
the focal plane, used for reconstructing the particle
trajectories, are followed by two pairs of segmented plastic
scintillators that provide the main trigger signal as well as the
time-of-flight information.  The time-of-flight resolution is $\sim
150$~ps $(\sigma)$.  For electron identification, a lead-glass shower
detector array together with a gas threshold \v Cerenkov is used in
order to distinguish between $e^-$ and $\pi^-$. For kaon
identification in the SOS, a silica aerogel detector (n=1.034)
provided $K^+/\pi^+$ discrimination while an acrylic \v Cerenkov
counter (n=1.49) was used for $K^+/p$ discrimination.  

Electroproduction processes involve the exchange of a virtual photon,
$\gamma^*$, between projectile and target. The spectrometer setting for
electron detection was kept fixed at an angle of 14.93${^\circ}$ during the
experiment, thereby holding the virtual photon flux constant
(cf.~Ref.~\cite{Zeidman:2001sh}).  The initial spectrometer angle of
the kaon arm was $13.40{^\circ}$. This angle was varied to measure angular
distributions with respect to the direction of $\gamma^*$. For the
$\gamma^*$, the invariant mass was $Q^2 = 0.35$~(GeV/c)$^2$, the virtual
photon momentum was $|\vec{q}|=1.77$ GeV/c and the total energy in the
photon-nucleon system was $W= 1.91$~GeV.  Electroproduction on light
nuclei was studied for three different angle settings with respect to
the initial kaon angle, $13.40{^\circ}$.  The corresponding angle between
the virtual photon, $ \gamma^*$ and the ejected kaon ($K$), are
$\theta_{\gamma^*K^+}^{\mathrm{lab}} \simeq 1.7{^\circ},~6{^\circ},~12{^\circ}$.
These correspond to increasing the momentum transfer to the hyperon
($\mid t\mid \simeq (0.12,\; 0.14,\; 0.23) \; \mathrm{GeV}^2$).  The
central spectrometer momenta were 1.29~GeV/c for the kaon arm and
1.57~GeV/c for the electron arm.

\section*{Data analysis}
\begin{figure}[phbt]
  \centering
  \includegraphics[width=12cm]{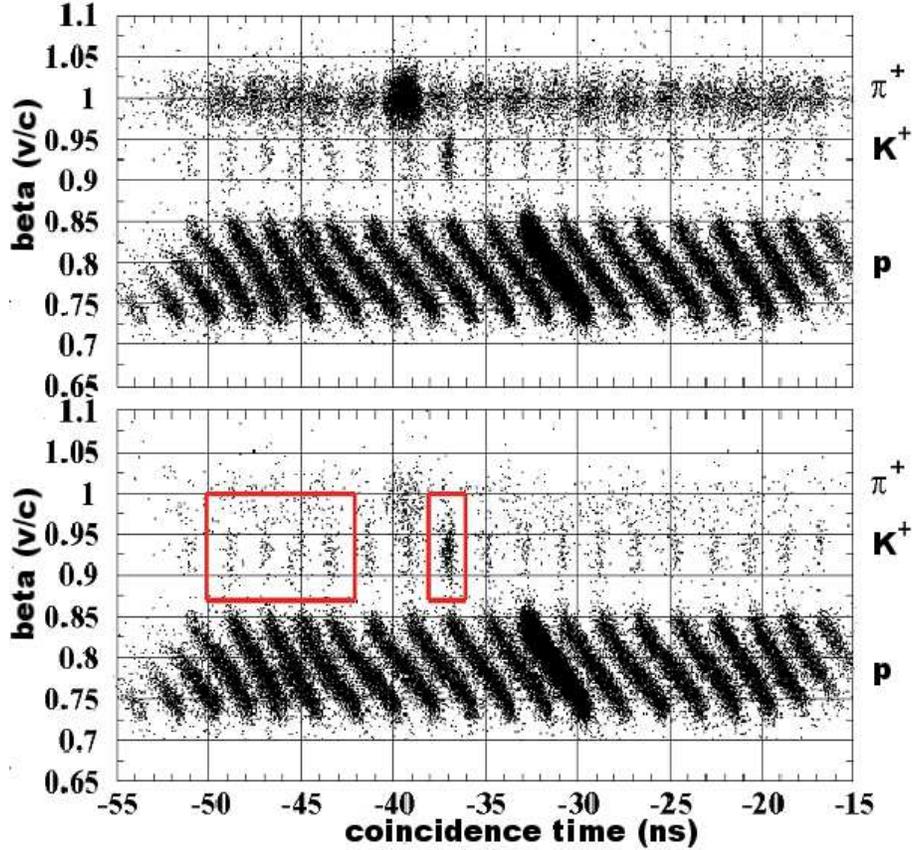}
  \caption{(Color online) Real and random events of $\beta_{K^+}$ versus the
    path-length corrected coincidence time measured by the SOS
    spectrometer. Visible bands correspond to protons (low
    velocities), kaons and pions (high velocities). The tilt of the
    pion and proton bands reflects that $\beta$ was calculated
    assuming the particle was a kaon. The effect of PID cuts, is shown
    in the bottom figure, where the fast pions were almost totally
    removed. The random events are determined by averaging over a
    number of random coincidence peaks as indicated by the large red
    box. These are to be subtracted from the small red box around the
    main coincidence peak.}
  \label{fig:coinnew}
\end{figure}

The essential element of the data analysis for the present work is a
clear identification of scattered electrons coincident with kaons
against a large background of pions and protons.
Figure~\ref{fig:coinnew} shows the measured hadron velocity in the SOS
versus the coincidence time between the two spectrometers. The latter
has been projected back to the target by using the kaon mass as
default. It thus represents the proper coincidence time only for
kaons, the particles of interest. Clearly visible is the 2-ns RF time
structure of the beam. The top panel shows the distributions before,
the bottom panel after applying an analysis cut on the aerogel \v
Cerenkov detector.  In-time electron-kaon coincidences are selected by
a cut on $\beta$ and coincidence time. The background from
uncorrelated $(e, K^+)$ pairs was subtracted using distributions from
out-of-time coincidences, a standard procedure for Jefferson Lab Hall
C experiments\cite{Gaskell:2001xr, Ambrozewicz:2004kn}.  Defining the
out-of-time window such that it does not include any in-time
coincidences of $(e, \pi)$ and $(e,p)$, this procedure also corrects
for any remaining pion and proton background in the in-time kaon
window.

\begin{figure}[htbp]
  \centering
  \includegraphics[width=0.8\textwidth]{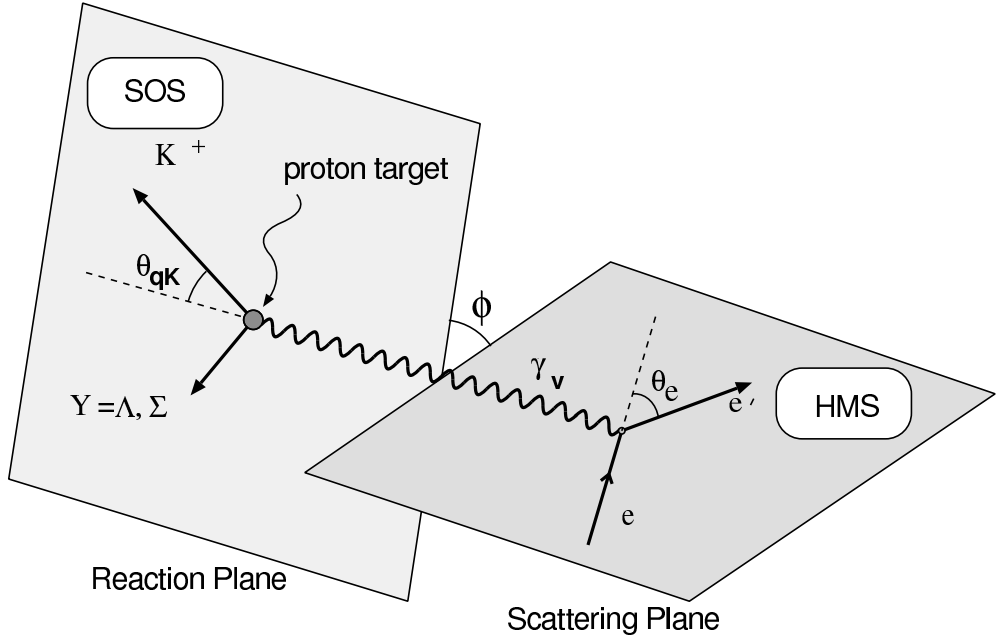}  
  \caption{The kinematics of kaon electroproduction:~the reaction (hadron)
    and scattering (lepton) planes are connected by the virtual photon
    which lies in both planes. The electron scattering angle is
    denoted by $\theta_e$, the kaon scattering angle between the kaon
    and the direction of the virtual photon is denoted by
    $\theta_{\gamma K}$. Typically for electroproduction experiments
    in Hall C of JLab, the ejected $K^+$ was detected by the
    SOS spectrometer in coincidence with the scattered $e'$, detected
    by the HMS spectrometer [from~\cite{Mohring:2002tr}].}
  \label{fig:electroproduction}
\end{figure}

Following Ref.~\cite{Knochlein:1995qz,Stijn:2002xy},
the notation of strangeness electroproduction may be introduced by
\begin{equation}
\label{eq:8}
p (p^{\mu}) + e(q^{\mu}_e) \to e^{\prime}(q^{\mu}_{e^{\prime}}) +  K(p^{\mu}_K) +
Y(p^{\mu}_Y) ,
\end{equation}
with the four-momenta $q^{\mu}_e = (E_e,\vec{q}_e)$, $q^{\mu}_{e^{\prime}} =
(E_{e^{\prime}},\vec{q}_{e^{\prime}})$ of the incoming and outgoing electron,  
$q^{\mu} = (\omega, \vec{q})$  
of the  virtual photon,  
$p^{\mu}_p = (E_p, -\vec{q})$, $p^{\mu}_K= (E_K, \vec{p}_K)$, $p^{\mu}_Y = (E_Y,-\vec{p}_K)$. The virtual photon is defined by the difference
of the four-vectors of the incoming and outgoing electron, $q^{\mu} =
q^{\mu}_e - q^{\mu}_{e^{\prime}}$. The kinematics are shown in
Fig.~\ref{fig:electroproduction}, where the lepton and hadron planes
are defined. The virtual photon connects both planes kinematically.

\begin{figure}[phbt]
  \centering
  \includegraphics[width=0.9\textwidth]{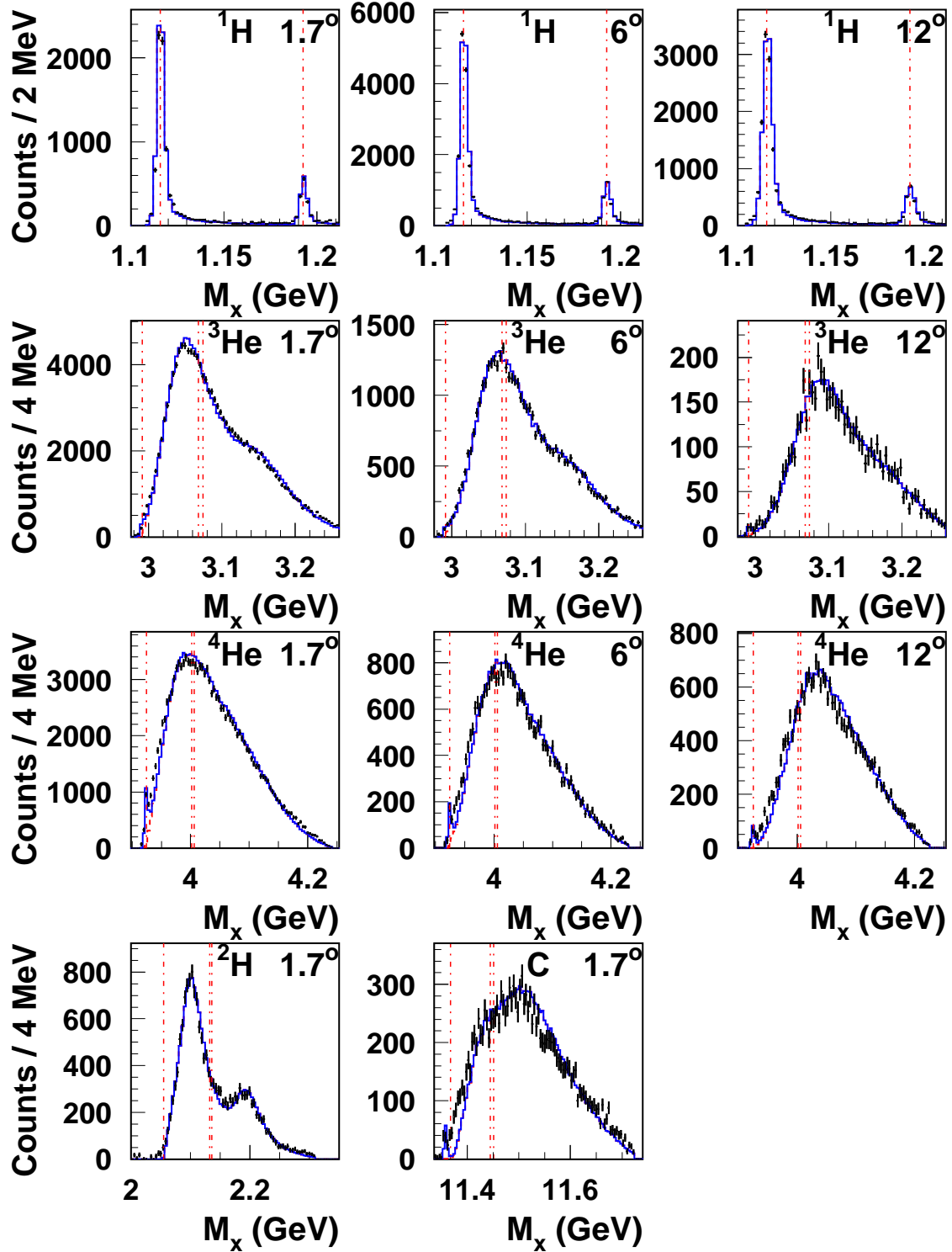}
  \caption{(Color online) Reconstructed missing mass spectra for all
    five targets at all kinematic settings as indicated.  For $^1H$,
    $^3$He, $^4$He, three kinematic settings were measured, whereas
    for $^2H$, and C targets only one kinematic setting
    ($1.7^{{^\circ}}$) was measured. The blue line represents the
    respective fit to each spectrum. }
  \label{fig:3x4}
\end{figure}

After proper electron and kaon identification, the measured momenta
(magnitude and direction with respect to the incoming beam) allow for
a full reconstruction of the missing energy and missing momentum of
the recoiling system:

The missing energy and missing momentum of the recoiling nucleons are
calculated \textit{viz.\ }
\begin{align}
E_X &= E_{e} - E_{e^{\prime}}+M_{\mathrm{targ}}-E_K = \omega + M_{\mathrm{targ}} - E_K, \\
\vec{P}_X &= \vec{q} - \vec{p}_K,
\end{align}
where $M_X = \sqrt{(E_X^2-|\vec{P}_X|^2)}$ is the missing mass,
$M_{\mathrm{targ}}$ denotes the target mass.  
The four-momentum transfer to the nucleons is given
by the Mandelstam variable $t$,
\begin{equation}
t = (q^{\mu}- p_K^{\mu})^2 = (\omega - E_K)^2 - |\vec{q}|^2 - |\vec{P}_K|^2 + 2|\vec{q}||\vec{P}_K|\cos{\theta_{pK}}\,.
\end{equation}
Final states of the $A(e,e'K)X$ reaction for $A=1,2,3,4,12$ are visible
in Fig.~\ref{fig:3x4}. 
The missing mass $M_X$ is calculated from the four momenta $q^{\mu}$ of
the virtual photon and the four momentum $p^{\mu}_K$ of the detected kaon,
\textit{viz.\ }
\begin{equation}
  \label{eq:1}
  M_X^2 = ( q^{\mu} + P_{\mathrm{targ}}^{\mu} - p_K^{\mu} )^2,
\end{equation}
where $P^{\mu}_{\mathrm{targ}} = (M_{\mathrm{targ}},0,0,0)$ is the
target four-momentum.

Missing mass distributions have been created for the in-time (e,K)
coincidences as well as a sample of the out-of-time coincidences; the
latter then were subtracted with the appropriate weight. For the
cryogenic targets, the background from the target cell walls was
determined by a measurement from an empty cell replica. Data from this
replica were subjected to the same analysis and subtracted from the
distributions.

Figure~\ref{fig:3x4} shows background subtracted missing mass
distributions for all four targets.  For the hydrogen target, the
missing mass distributions allow for an unambiguous identification of
the electroproduced hyperon, either a $\Lambda$ or a $\Sigma^0$. The
well known masses of these two hyperons also serve as an absolute mass
calibration with an accuracy of better than 2 MeV.

On the deuterium target, the two distributions are significantly
broadened because of the presence of a nucleon spectator and the Fermi
motion of the target nucleons. Furthermore, the $\Sigma$ distribution
now is comprised of two possible final states, either a $\Sigma^0 n$ or a
$\Sigma^- p$; the latter from the reaction with a neutron inside the
target. Since the mass difference between
$\Sigma^0$ and $\Sigma^-$ is small compared to the width of the distributions,
these two final states are completely unresolved. In
Fig.~\ref{fig:3x4} it is also obvious that the radiative tail from the
$\Lambda$ distribution contributes significantly to the strength observed
in the $\Sigma$  mass region.  For increasing $A$, the peaks associated
with $\Lambda$ and $\Sigma$ hyperons further broaden. Whereas for $^3$He
a small shoulder associated with $\Sigma$ is still visible,
only an indistinct broad distribution remains for the $^4$He target.

This challenges any extraction of the underlying three reaction
channels $\gamma^*+p \to \Lambda +K^+$, $\gamma^* +p \to \Sigma^0+K^+$,
and $\gamma^*+n \to \Sigma^- +K^+$. The following section will
describe an attempt to disentangle the three reaction channels by
means of a Monte Carlo simulation that models the spectrometer
acceptances as well as the reaction mechanism.

The electroproduction cross section may be written as follows:
\begin{equation}
\frac{d^5\sigma }{dE_{e^{\prime}} d\Omega_{e^{\prime}} d\Omega_K} =
\Gamma \frac{d^2\sigma}{d\Omega_K}
\end{equation}
where $\Gamma$ denotes the virtual photon flux factor:
\begin{equation}\label{eq:22}
\Gamma = \frac{\alpha}{2\pi^2}\frac{E_{e^{\prime}}}{E_e}\frac{1}{Q^2}\frac{W^2-M^2}{M}\frac{1}{1-\varepsilon},
\end{equation}
where $\alpha$ is the fine structure constant and $\varepsilon$ is the
longitudinal polarization of the virtual photon,
\begin{equation}
\varepsilon = \left( 1 + 2\frac{|\vec{q}|^2}{Q^2}\tan^2(\theta_e/2)\right)^{-1}.
\end{equation} 
The total energy in the
virtual photon--target center is given by
$W^2=s=(q^{\mu}+p^{\mu}_{\mathrm{target}})^2$ and can be expressed in
the laboratory reference frame by $W^2 = M^2 +2M\omega -Q^2$. To
facilitate comparison with the scattering on the proton, both for
calculating $W$ as well as in Eq.~(\ref{eq:22}) $M$ is taken to be the
nucleon mass for all targets discussed here.

The ${}^1\mathrm{H}(e,e'K^+)X$ data was used to provide consistent
normalization data as well as to test available isobar models and to
develop a global model that would describe the data. While reasonable
agreement was found with the Saclay-Lyon model~\cite{David:1995pi},
the best description of the data within the kinematic range of this
experiment was achieved by a dedicated simple model. This model had
already been developed for the first experimental run on $A=1,2$
targets~\cite{Cha:2000xy}.  
Unlike the Saclay-Lyon model it is not based on separated response
functions. Instead the unpolarized two-fold center of mass cross 
section is modeled and taken as input for the simulations, which
then provides a five-fold laboratory cross section as output.

The model describes the unpolarized
differential cross section for ${}^1\mathrm{H}(e,e'K^+)\Lambda$ by a
factorization ansatz of four kinematic variables:
\begin{equation}
  \left.\frac{d^2\sigma}{d\Omega}\right|_{\Lambda}\left(Q^2,W,t,\phi\right)  = f(Q^2)\times N\cdot g(W)h(t)i(\phi),\label{eq:2}
\end{equation}
with a normalization constant $N=5.4724$ and the four functions
\begin{align}
  f(Q^2) &= \mathrm{constant} = c_1^f,\label{eq:3}\\
  g(W) &= c^g_1\frac{P_K^{\mathrm{cm}}}{(W^2-M_p^2)W}+\nonumber\\
       &+ c^g_2\frac{W^2}{c^g_3W^2+(W^2-1.72^2)^2},\label{eq:4}\\
  h(t_{\mathrm{min}}-t)&= \exp(c^h_1(t_{\mathrm{min}}-t)),\label{eq:5}\\
  i(\phi)&=c_1^i+c_2^i\cos(\phi)+c_3^i\cos(2\phi)\label{eq:6}.
\end{align}
The $c_{1,2,3}^{f,g,h,i}$ are parameters which are determined through
a fit to the data taken during the first experimental
run~\cite{Cha:2000xy,Reinhold:2001zm}. These parameters are given in
Table~\ref{tab:jinparams}.

The functional form of the t dependence in Eq.~\eqref{eq:5} has been
taken from an earlier work by Brauel et al.~\cite{Brauel:1979zk},
while the $\phi$ dependence was studied during the first run of the
experiment~\cite{Cha:2000xy}.  Equation~\eqref{eq:4} shows that the
dependence on the total photon energy $W$ is composed of a phase space
factor and a Breit-Wigner resonance.  The observed $Q^2$ dependence is
very weak and it is set to a constant.

For the electroproduction of $\Sigma^0$ hyperons,
${}^1\mathrm{H}(e,e'K^+)\Sigma^0$, only a single, energy dependent
phase space factor is used. Following \cite{Bebek:1976qg} we obtain
\begin{equation}
  \label{eq:10}
  \left.\frac{d^2\sigma}{d\Omega}\right|_{\Sigma}\left(W\right)=c_1\frac{P_K^{\mathrm{cm}}}{(W^2-M_p^2)W};\quad c_1=1.32 \,\mathrm{GeV}^2 \mu\mathrm{b/sr}
\end{equation}
where the constant $c_1$ was determined by
Koltenuk~\cite{Koltenuk:1999jv}.

Unlike hydrogen, the missing mass distributions for deuterium and the
other nuclear targets do not show two clearly separable peaks, cf.\
Fig.~\ref{fig:3x4}, as discussed above.  To extract information on the
quasifree $\Sigma^0$ as well as $\Sigma^-$ production, one has to rely
on assumptions about the nuclear dependence of the $\Sigma^0$.  In
this analysis, we determine the ratio of $\Lambda$ to $\Sigma$
production for hydrogen and then keep this ratio fixed in the proton
model that enters into the simulation for the nuclear cross section.
Nuclear effects thus contribute to the systematic uncertainties
(cf~\cite{Abbott:1998tq, Kohri:2006yx}). If such an assumption is not
made, only a combined $\Sigma$ contribution may be deduced, as
in~\cite{Boyarski:1970yc, Quinn:1979zp}.

The data shown in Fig.~\ref{fig:3x4} were compared with a dedicated
Monte Carlo simulation that modeled the spectrometer optics and
acceptance, kaon decay, small angle scattering, energy loss and
radiative corrections~\cite{Ent:2001hm, Mohring:2002tr}.  The process
of extracting the respective cross sections described in
detail in~\cite{Ambrozewicz:2004kn, Gaskell:2001xr}, relies upon
a ratio of the measured yield from experiment, $Y_{\mathrm{exp}}$,
normalized to a simulated yield from the above mentioned Monte Carlo
simulation, $Y_{\mathrm{MC}}$, which is used as a scale factor for the
model cross section used in the Monte Carlo, \textit{viz.\ }
\begin{equation}
  \label{eq:7}
  \frac{d^2\sigma}{d\Omega} = \frac{Y_{\mathrm{exp}}}{Y_{MC}}\cdot \left.\frac{d^2\sigma}{d\Omega}\right|_{\mathrm{model}}.
\end{equation}
This approach is also known as the method of correction factors,
cf.~\cite{Cowan:1998ji}.  For $A=2,3,4,12$ the $A(e,e'K^+)X$ process
was modeled as quasifree scattering on target nucleons inside the
target.  Since to the best knowledge of the authors no dedicated
models are available for the electroproduction on these nuclei, the
elementary cross section model eqs.~\eqref{eq:2}--\eqref{eq:6} for
$\Lambda$ and eq.~\eqref{eq:10} for $\Sigma$, are used. The respective
cross sections are multiplied by the number of protons, $Z$, or
neutrons, $N$, respectively.  Since no separate model for the
production on the neutron is available, we use the
model~\eqref{eq:10} for both $\Sigma^0$ as well as $\Sigma^-$. The
model is convolved with spectral functions~\cite{Benhar:1994hw} for
the respective target nucleus.
 
The spectral functions provide the four-momenta of the target nucleons
inside the target. For the $A=2$ case, deuteron momentum distributions
taken from either the Bonn potential \cite{Machleidt:1987hj} or the
Av18 potential \cite{Wiringa1995} gave essentially identical results.
Obviously neither of these models incorporate any possible in-medium
behavior of the nucleons inside the target nor final state interaction
as will be discussed below.  For the nuclear targets, final state
interactions in the vicinity of the respective quasifree thresholds
are taken into account using an effective range
approximation~\cite{Gillespie:1964xy}.

The final state interaction of the hyperon with the remaining target
nucleon has to be taken into account, whereas the kaon nucleon final
state interaction is small; the $\Lambda N$ total cross section is
more than two orders of magnitude larger than the $K^+N$ total cross
section\footnote{http://pdg.lbl.gov/2006/\mbox{hadronic-xsections}/hadron.html}.  We use an effective range approximation
(ERA), by which the modeled cross section is modified by an
enhancement factor $I$ (cf.\ Watson and
Migdal~\cite{Watson:1952ji,Migdal:1955xy}),
\begin{equation}\label{eq:11}
\sigma_K^{YN\,\mathrm{FSI}} = I\sigma_K =
\frac{1}{|J_l(k_{\mathrm{rel}})|^2}\cdot\sigma_K,
\end{equation}
in terms of the complex Jost function $J_l$ for the $l$th partial
wave. $k_{\mathrm{rel}}$ is the relative momentum between the hyperon
and the nucleon (see also chapters 12 and 14 of~\cite{Newton:1982qc}).
A hyperon--nucleon ($YN$) potential $V$ is used to describe the final
state interaction, for which only the s-wave part is taken into
account.  The s-wave Jost function may then be written as
\begin{equation}\label{eq:12}
J(k_{\mathrm{rel}}) = \frac{k_{\mathrm{rel}}-i\beta}{k_{\mathrm{rel}}-i\alpha},
\end{equation} 
where $\alpha$ and $\beta$ are determined from the scattering length $a$ and effective range $r_e$ of the hyperon-nucleon potential \textit{viz.\ }
\begin{equation}\label{eq:13}
\frac{1}{2}r_e(\alpha - \beta) = 1\,,\quad \frac{1}{2} r_e \alpha\beta = -\frac{1}{a}\,.
\end{equation}
In this ansatz there are no free parameters, the magnitude of the
enhancement factor is fully determined by the effective range $r_e$
and the corresponding scattering length $a$, both being parameters of
the hyperon-nucleon potential chosen.  For the $A=2$ targets, the full
Jost function ansatz gave a less satisfactory description of the data
than for the helium targets. An even simpler approach for an ERA,
studied in~\cite{Cha:2000xy} and following a prescription described in
reference~\cite{Li:1991qa} was used. The s-wave phase shift $\delta_0$
is calculated via the Bethe formula and the enhancement factor is
given by
\begin{gather}
  k_{\mathrm{rel}}\cot \delta_0 = -\frac{1}{a}+ 0.5 r_e k_{\mathrm{rel}}^2\quad
  I = \left(\frac{\sin(\delta_0 +k_{\mathrm{rel}}r)}{\sin{k_{\mathrm{rel}}r}}\right)^2.
\end{gather}

\begin{figure}[hbtp]
  \centering
  \includegraphics[width=12cm]{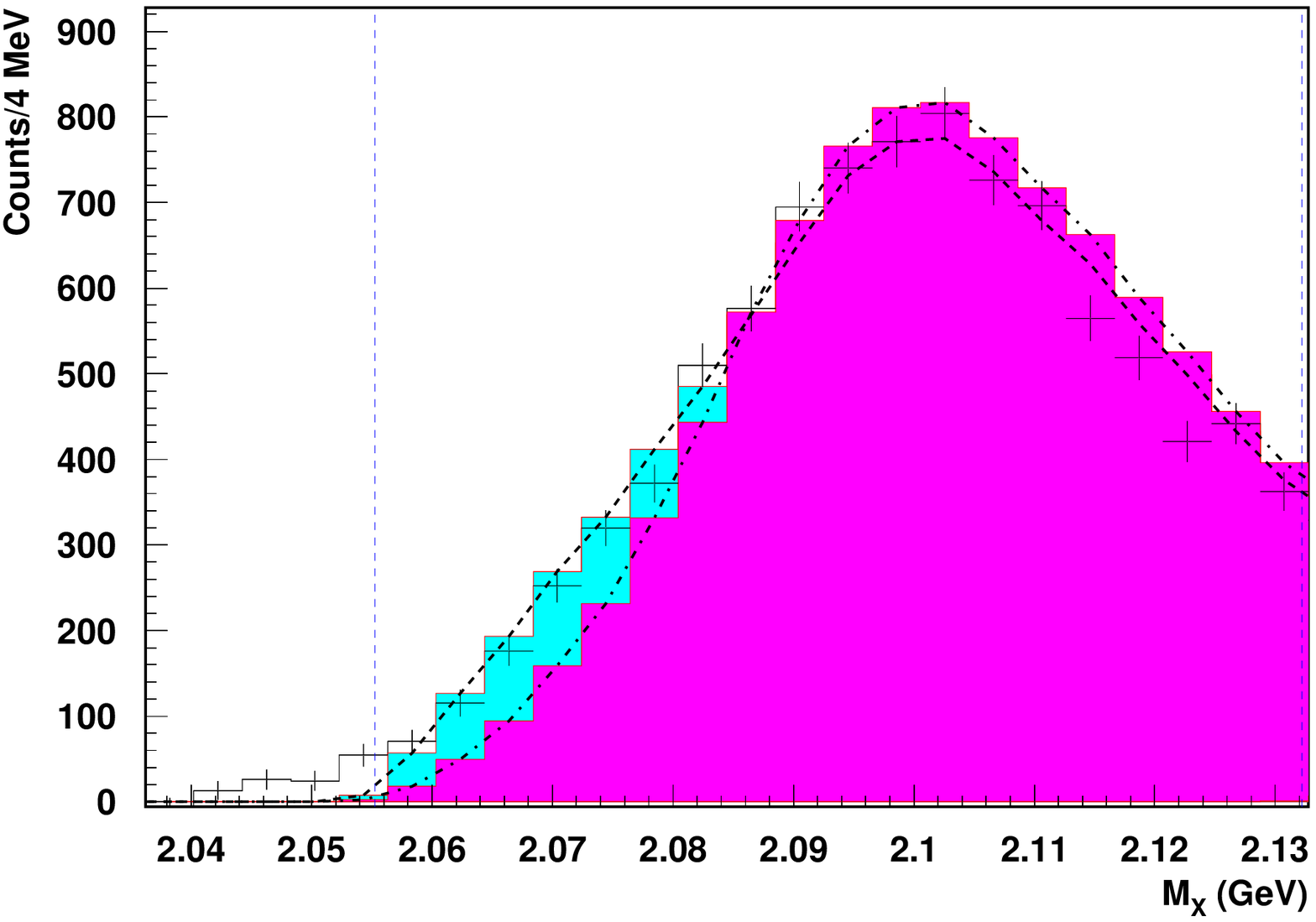}
  \includegraphics[width=12cm]{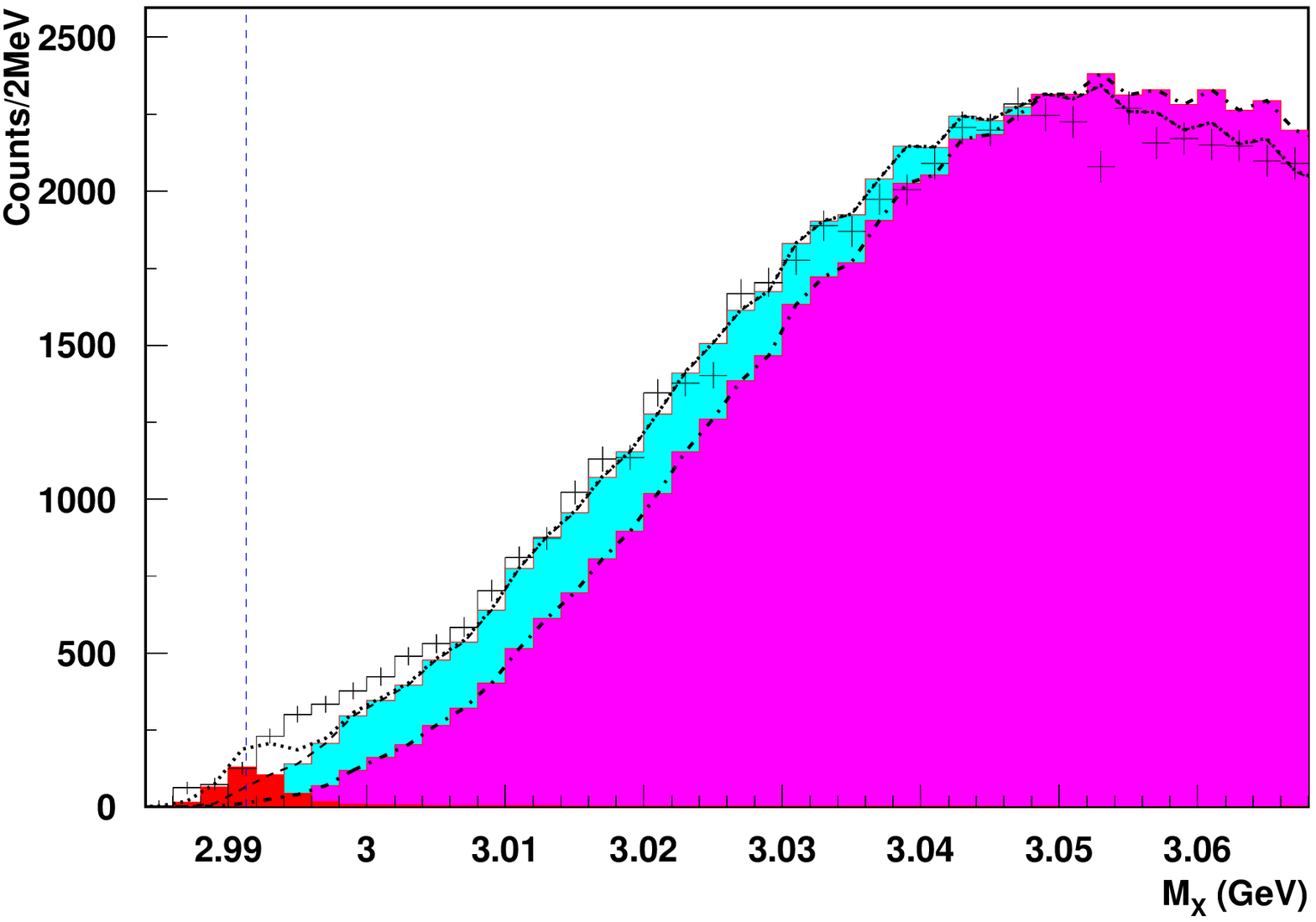}
  \caption{(Color online) The effects of including FSI for the fits to
    the data on $^2$H (upper panel) and $^3$He (lower panel) in the
    low-mass $\Lambda$ region. The fitted $\Lambda$ contribution
    without FSI is given by the dark color, dash-dotted line.
    $\Lambda$ contributions including FSI are given by the light-blue,
    dashed line. For $^3$He, the $^3_{\Lambda}$H bound state is shown
    in red. The total fit (sum of all contributions) is given by the
    dotted line. The vertical dashed line denotes the threshold for
    $\Lambda$ production on $^2$ H, $^3$He, respectively.}
  \label{fig:fsivis}
\end{figure}

For the helium targets, however, the full Jost function ansatz gave
much better results. For the data sets presented in this
paper, we use the Nijmegen 97f $YN$ potential~\cite{Rijken:1998yy}, with
scattering lengths $a$ taken from~\cite{Rijken:1998yy} and effective
ranges of $r_e$ taken from the Nijmegen 89\cite{Maessen:1989sx}, since
Ref.\ \cite{Rijken:1998yy} does not provide these parameters. In all
cases and for every hyperon--nucleon potential tested, the singlet
values for $a$ and $r_e$ gave more satisfactory results than triplet
values.  For the $\Sigma$ hyperons, the Nijmegen 97f and the
J\"ulich~A also provide $a$ and $r_e$ for the $\Sigma N$ interaction.
Using these values, an enhancement factor due to $\Sigma N$ final
state interaction was introduced. However, the fits to the data were
more strongly influenced by the $\Lambda N$ final state interaction.
In Fig.~\ref{fig:fsivis} we show the effect of applying final state
interaction in an ERA to our model in the low-mass $\Lambda$ region.

\begin{figure}[phbt]
  \centering
  \includegraphics[width=0.9\textwidth]{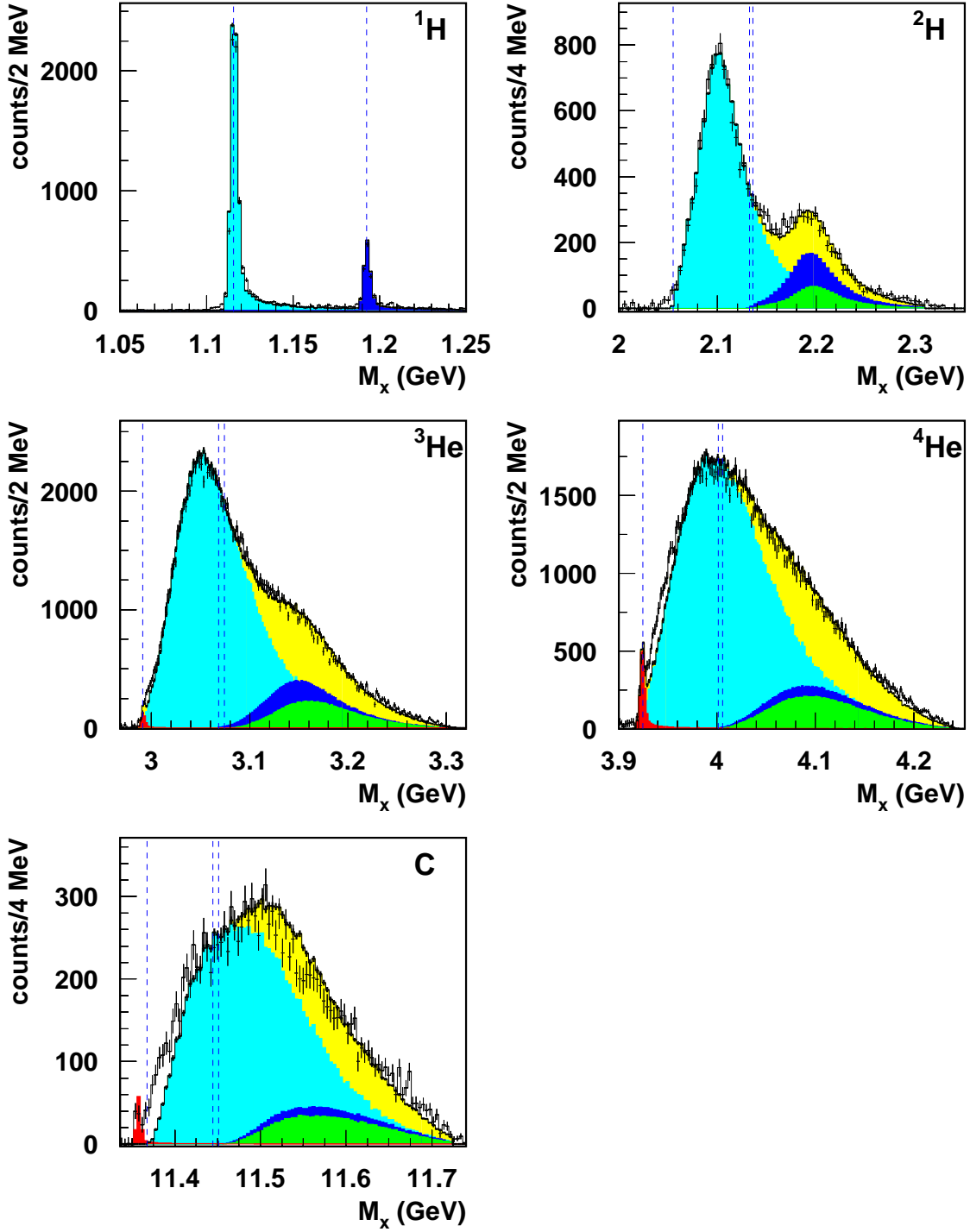}
  \caption{(Color Online) Reconstructed missing mass spectra for six
    targets at one kinematic setting ($\langle
    1.7^{{^\circ}}\rangle$). The lowest lying thresholds for quasifree production of
    $\Lambda$ and $\Sigma$ hyperons on each targets are indicated by
    the dot-dashed vertical lines. For hydrogen, these lines
    correspond to the pole masses of the $\Lambda$ and $\Sigma^0$
    hyperons, respectively. Simulated quasifree reactions
    A$(e,e'K^+)$Y are indicated by colors: $Y=\Lambda$ (lightblue),
    $Y=\Sigma^0$ (blue), $Y=\Sigma^-$ (green), bound states
    $^3_{\Lambda}$H, $^4_{\Lambda}$H, $^{12}_{\Lambda}$B (red), sum
    of all simulated contributions (yellow).}
  \label{fig:fit6_0}
\end{figure}

In Table~\ref{tab:fsi} we show the influence of the FSI on the
simulated missing mass yields. The simulated missing mass is weighted
by the respective model cross section. If the cross section is
multiplied by an enhancement factor, the missing mass spectra is
influenced. Table~\ref{tab:fsi} gives the ratio of the integrated
yields $Y_{\mathrm{FSI}}/Y_{\mathrm{no\_FSI}}$ for missing mass
distributions (cf.\ Figs.~\ref{fig:3x4} and~\ref{fig:fit6_0}) with or
without FSI for the model~\eqref{eq:2} discussed above.  Choosing a
different cross section model would change these values only by
1\%--3\%.  cross section models. Also, different final state
interaction models (e.g.\ Nijmegen 97f, J\"ulich A) do not change the
yield ratio by more than 3\%.

For the helium-3 and helium-4 target nuclei (and also for carbon), the
analysis was performed analogously to the $A=2$ case.  However, the
electroproduction of strangeness on helium targets (and on carbon,
though with a rather poor statistics) triggers two investigations: the
quasifree production of open strangeness on the light nuclear target
as well as the production of bound hypernuclear states.  The missing
mass distributions for these targets are shown in Figs.~\ref{fig:3x4}
and~\ref{fig:fit6_0}. It is obvious from both figures that the
investigation of the quasifree reactions on the one hand and
structures near the respective thresholds for quasifree production do
not completely decouple due to the limited mass resolution of the
missing mass distributions. Therefore the quasifree distribution and
the coherent distribution overlap.

The following describes the extraction of the cross section: For the
${}^1\mathrm{H}(e,e'K^+)$ data, we fit the missing mass spectra
$M^{\mathrm{data}}$ with the following
ansatz:
\begin{gather} \label{eq:14} M^{\mathrm{data}}(\mathrm{H}) =
  f_{\mathrm{H},{\Lambda}}\cdot
  M_{\Lambda}^{\mathrm{model}}(\mathrm{H}) +
  f_{\mathrm{H},{\Sigma^0}}\cdot
  M_{\Sigma^0}^{\mathrm{model}}(\mathrm{H}),
\end{gather}
with two free fit parameters $f_{\mathrm{H},{\Lambda}}$ and
$f_{\mathrm{H},{\Sigma^0}}$ for the simulated missing mass
distributions $M_{\Lambda,\Sigma^0}^{\mathrm{model}}$. Once these two parameters are obtained,
the cross section in the laboratory may be obtained by evaluating the
model cross section for the simulation at the specific kinematic
conditions of the experiment, as stated above. These two model cross
sections are then multiplied by the respective fit parameters obtained 
in~(\ref{eq:14}). Moreover, we define the important ratio of the fit parameters  
\begin{equation}
  \label{eq:15}
  R_{\Lambda \Sigma^0} = \frac{f_{\mathrm{H},{\Lambda}}}{f_{\mathrm{H},{\Sigma^0}}}.
\end{equation}
For targets with $A\geq2$ Eq.~(\ref{eq:14}) has to be modified to
incorporate the possible conversion of a target neutron into a
$\Sigma^-$  hyperon as follows:
\begin{widetext}
  \begin{equation} \label{eq:17} M^{\mathrm{data}}(A) =
    f_{A,{\Lambda}}\cdot M_{\Lambda}^{\mathrm{model}}(A) +
    f_{A,{\Sigma^0}}\cdot M_{\Sigma^0}^{\mathrm{model}}(A) +
    f_{A,{\Sigma^-}}\cdot M_{\Sigma^-}^{\mathrm{model}}(A).
  \end{equation}
\end{widetext}
Here the simulated missing mass distributions $M_{Y}^{\mathrm{model}}(A)$,
$Y=\Lambda, \Sigma^0, \Sigma^-$ include both the respective
model cross section and the respective enhancement factors $I_{Y}(A)$due
to final state interaction.   
The respective cross sections are given by 
\begin{equation}\label{eq:16}
\sigma_{Y}(A) = f_{A,Y}\cdot I_{Y}(A)\cdot
\sigma^{\mathrm{model}}_{Y}(A).
\end{equation}
In the following, if not explicitly
stated otherwise, it is assumed that the model cross section
$\sigma_Y^{\mathrm{model}}(A)$ themselves do not include final state
interaction. Enhancements of the model cross sections due to final
state interaction are described by enhancement factors $I_Y(A)$. 

Eq.~(\ref{eq:17}) poses a fitting problem with three free fit
parameters $f_{Y}(A)$ for which this experiment is not able to 
distinguish directly the contributions of either $\Sigma$ hyperon.
Thus for targets with $A\geq2$, it is assumed that this ratio
\eqref{eq:15} is the same for the bound protons in the respective
nucleus, i.e.\ 
\begin{equation}
  \label{eq:15a}
  R_{\Lambda \Sigma^0} =
  \frac{f_{\mathrm{H},{\Lambda}}}{f_{\mathrm{H},{\Sigma^0}}} =
  \frac{f_{\mathrm{A},{\Lambda}}}{f_{\mathrm{A},{\Sigma^0}}} \,, 
  f_{A,\Sigma^0}= \frac{f_{A,\Lambda}}{R_{\Lambda\Sigma_0}(\mathrm{H})}.
\end{equation}
Instead of fitting $f_{A,{\Sigma^0}}$, this parameter is calculated from
the fitted $f_{A,{\Lambda}}$, using the results from the
previous fit to the hydrogen data,  

With $f_{A\Sigma^0}$ determined via~(\ref{eq:15a}),~(\ref{eq:17})
reduces to a fitting problem with only two free parameters.

For $^3$He, $^4$He and $^{12}$C  there is one additional parameter 
to be taken into account. These missing mass spectra show $\Lambda$ 
bound states for the respective nuclear target. 
For $^4$He, a $^4_{\Lambda}$H bound state is clearly visible for all
three kinematic setting just below the $^3$H-$\Lambda$ threshold of
3.925~MeV (cf.\ Figs.~\ref{fig:3x4} and~\ref{fig:fit6_0}). For
$^3$He, just below the $^2$H-$\Lambda$ threshold of 2.993~MeV, the
$^3_{\Lambda}$H bound state is barely visible as a weak shoulder for
$1.7{^\circ}$, but clearly present for $6{^\circ}$ and $12{^\circ}$ (cf.\
Fig.~\ref{fig:3x4}). For carbon, the $^{12}_{\phantom{1}\Lambda}$B
bound state is
clearly visible in the respective missing mass spectrum.
The fits to the respective bound
states for the helium targets and carbon do include one extra term for
the bound state to be fitted. This extra term is not shown in
Eq.~(\ref{eq:17}) -- it however contributes only over very narrow
ranges of the fit and does not cause ambiguities in the procedure.

In the next section we focus on the extraction of the quasifree cross
sections, angular distributions, and nuclear dependence, for the
respective targets. 

\section*{Results and discussion}
The measurement presented in this work provides data for targets with
$A=1-4$, and carbon.  The fivefold differential cross sections,
$d^5\sigma$, as well as the twofold center of mass differential cross
section $d^2\sigma$ per nucleus are given in
Table~\ref{tab:values}. Unlike our previous paper on
hypernuclear bound states \cite{Dohrmann:2004xy}, these cross
sections have not been normalized to the number of contributing
nucleons $n$ (e.g.\ $^3$He: $n_{\Lambda}=n_{\Sigma^0}=2$,
$n_{\Sigma^-} = 1$; $^4$He:
$n_{\Lambda}=n_{\Sigma^0}=n_{\Sigma^-}=2$).

We chose a binned maximum likelihood method (cf.\
Ref.~\cite{Barlow:1993dm}) for fitting the simulated distributions to
the data. This procedure was already successfully used in another
electroproduction experiment using the same equipment
(cf.~\cite{Ambrozewicz:2004kn}).  The fits were not constrained to fit
the data only in specific regions of $M_X$.  The binning of the
respective missing mass distributions was chosen between 2-4 MeV and
had no noticeable effect upon the cross section extraction.
  
The angular distributions were restricted to a common range covered in
azimuthal angle ($180\pm 24{^\circ}$). For the settings with near
parallel kinematics, $1.7{^\circ}$, however, the full azimuth was
accessible.  The uncertainties given in Table~\ref{tab:values} reflect
statistical and fitting uncertainties. In the following, we discuss
systematic uncertainties to be added to the uncertainties in
Table~\ref{tab:values}.  These uncertainties are tabulated in
Table~\ref{tab:certain}.  Correlated systematic uncertainties due to
yield corrections, including efficiency corrections, dead times and
event losses are $\sim 3\%$, while uncorrelated uncertainties,
including time-of-flight determination ($\sim 2\%$), particle
identification ($\sim 2\%$), absorption of kaons in the spectrometer
and target material ($\sim 3\%$), and kaon decay ($\sim 3\%$) amount,
in total, to $\sim 5\%$ (cf.~\cite{Uzzle2002}), thereby yielding a
combined uncertainty of $\sim 6 \%$ from these sources. Uncertainties
due to the analysis approach will be discussed below.
 
For the extraction, separate $M_X$ distributions were generated for
quasifree production of $\Lambda$, $\Sigma^0$, and $\Sigma^-$
hyperons, and the sum of these spectra was fitted to the total kaon
$M_X$ spectrum using a maximum likelihood fit. The fit parameters
$f_A$ and $f_H$ (cf.\ (\ref{eq:14}) - (\ref{eq:15a})) were roughly of
order unity.  For $A=3$ and $A=4$, bound state contributions for
$^{3,4}_{\Lambda}$H, also included, were discussed in
Ref.~\cite{Dohrmann:2004xy}.  For carbon, however, the
$^{12}_{\Lambda}$B bound state is bound so deeply that omitting it
from the fits changes the respective cross sections by less than
0.3\%. We estimate the laboratory cross section for the
$^{12}_{\Lambda}$B to be on the order of $\sigma_{\mathrm{lab}} \sim
(.9\pm.2\,(\mathrm{stat}))$ nb/GeV/sr$^2$, $\sigma_{\mathrm{cm}}
\sim (17.8 \pm 4.5\,(\mathrm{stat}))$ nb/sr, where both cross sections
have been divided by $n_p=6$. We note however, that we do not resolve
ground or excited states of $^{12}_{\phantom{1}\Lambda}$B, as were
resolved in other experiments~\cite{Miyoshi:2002ts}, such that our
cross section estimate represents an integral value only.

The uncertainties of the cross section determination of
$\Sigma^0$ are tied to those of $\Lambda$, since the ratio of
$\Sigma^0$ to $\Lambda$ production is fixed to the hydrogen results.
However, any deviation from this assumption will result in large
uncertainties on the $\Sigma^-$ cross section extracted from nuclei.

Alternatively, we include a combined cross section for $\Sigma^0$ and
$\Sigma^-$, the sum of the extracted cross sections for both $\Sigma$
hyperons. We extracted the combined $\Sigma$ cross section from a unconstrained
fit of just two quasifree distributions for $\Lambda$ and $\Sigma$ to
the respective data for all targets. For the combined $\Sigma$
analysis, results agree with the main analysis within uncertainties
( $\leq 3\%$ for $\Lambda$, $\leq 10\%$ for $\Sigma$).

\begin{figure*}[htbp] 
  \begin{minipage}[t]{7.5cm}
    \includegraphics[width=\textwidth]{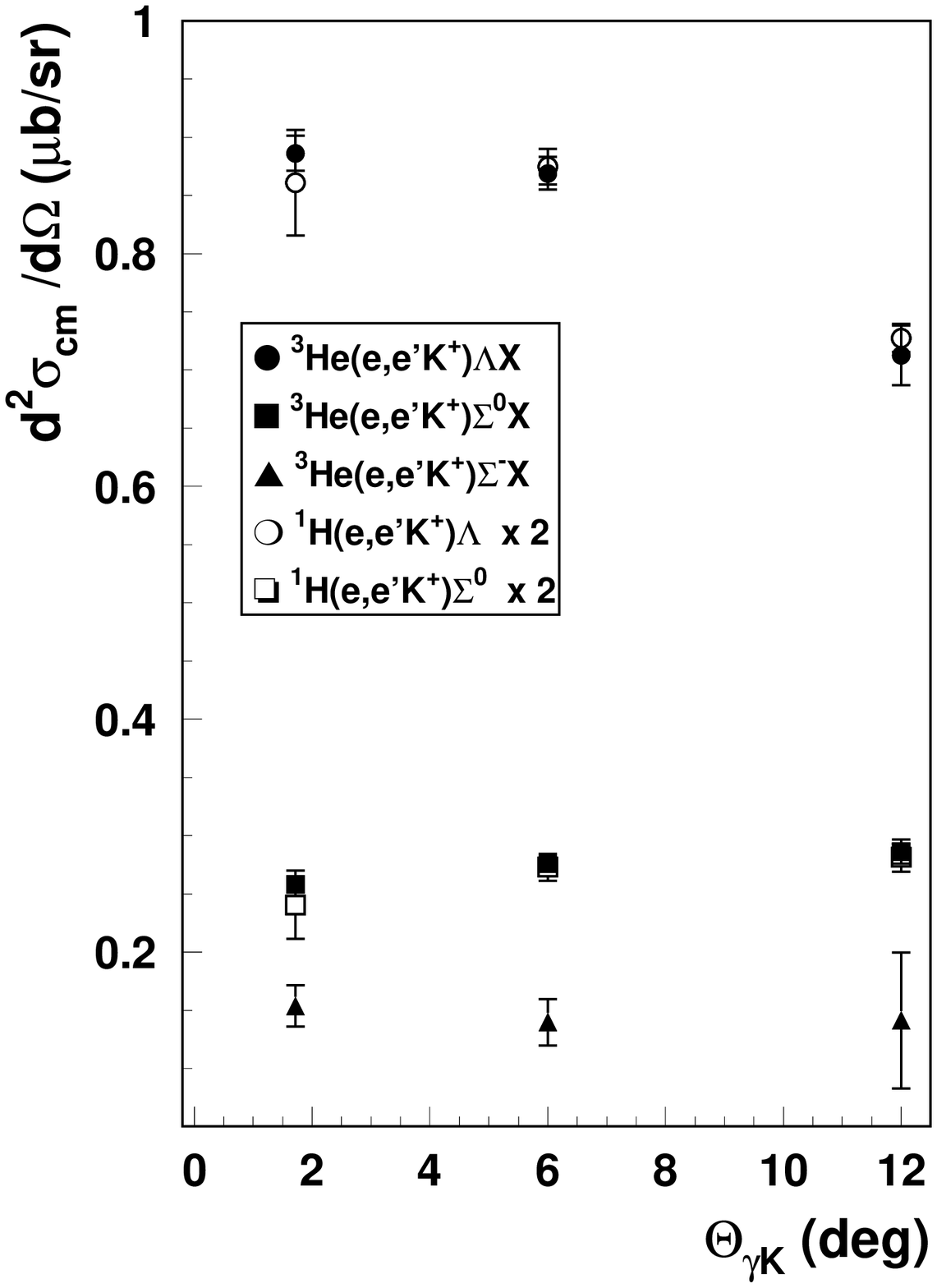}
    \caption{Comparison of the nuclear cross section for quasifree $\Lambda$,
      $\Sigma^0$ and $\Sigma^-$ production on $^3$He targets. For
      comparison, the respective quasifree distribution on the proton
      are shown by open symbols. These points have been scaled by a
      factor of 2 for better comparison.}
\label{fig:he3angdis}
\end{minipage} 
\hfill
\begin{minipage}[t]{7.5cm}
    \includegraphics[width=\textwidth]{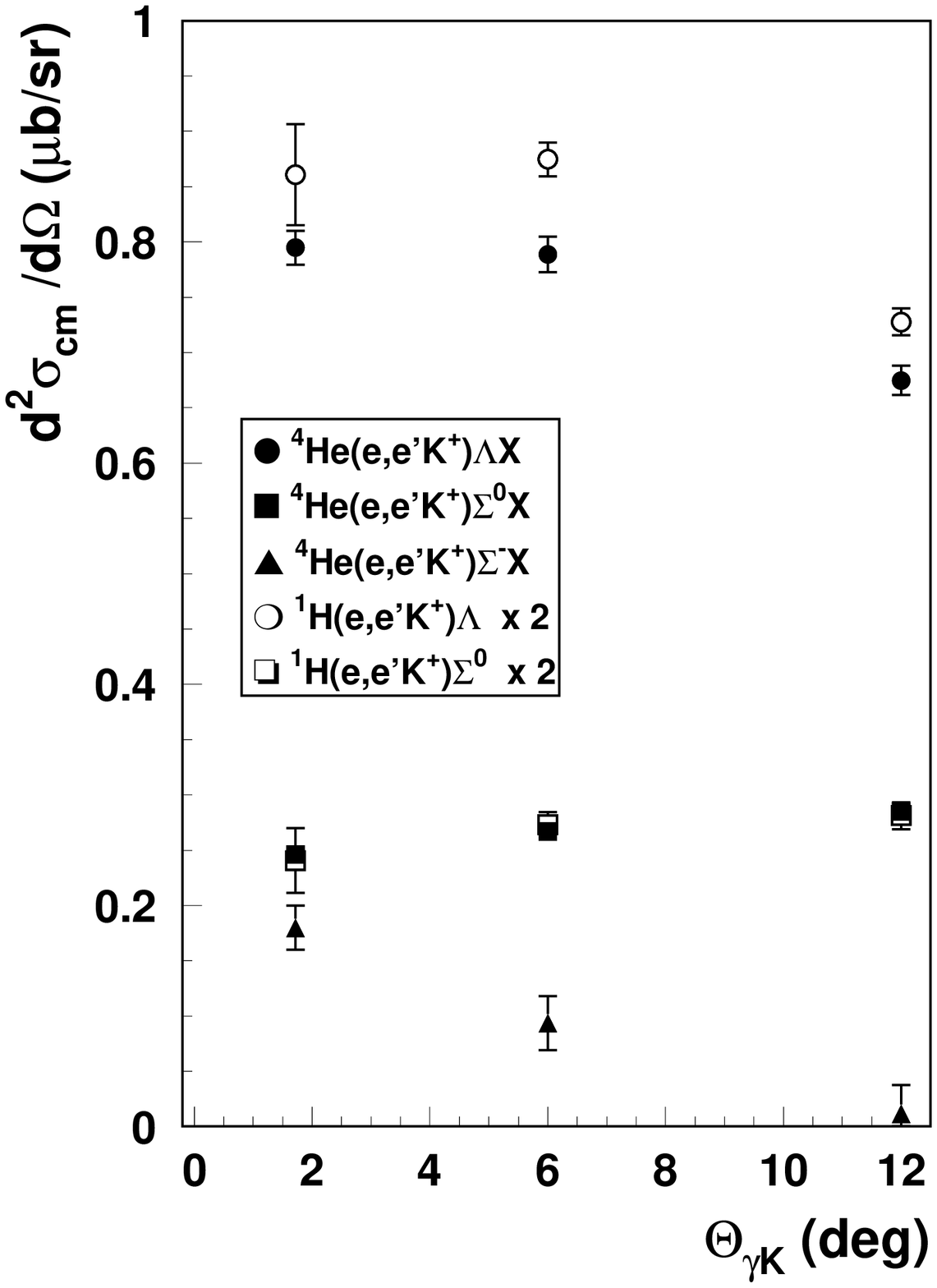}
    \caption{Comparison of the nuclear cross section for quasifree $\Lambda$,
      $\Sigma^0$ and $\Sigma^-$ production on $^4$He targets. The
      respective quasifree distributions on the proton are shown by
      open symbols. These points have been scaled by a factor of 2 for
      better comparison.}
\label{fig:he4angdis}
\end{minipage}
\end{figure*}

Figures~\ref{fig:he3angdis} and~\ref{fig:he4angdis} display the cross
sections for all three hyperons for ${}^{3,4}$He in the center of mass
system. For comparison, the quasifree distributions from hydrogen are
displayed as open symbols. For convenience, the hydrogen values have
been scaled by a factor of two.  In general, the
distributions are similar and seem to be strongly imprinted by the
underlying kinematics. While the angular distributions for the $\Lambda$
hyperon drop with increasing $\theta_{\mathrm{lab}}$, the $\Sigma^0$
distributions stay nearly flat. This is also observed for $^3$He and
$^4$He. Considerably different are the $\Sigma^-$ distributions for
the respective hyperons. For $^3$He, the $\Sigma^-$ angular
distribution does not show any strong dependence on the angle, similar
to the $\Sigma^0$ distribution. For $^4$He, however, the $\Sigma^-$
distribution drops significantly with angle.  With increasing angle,
the remaining strength seems to be exhausted by $\Lambda$ and
$\Sigma^0$ alone, so that the $\Sigma^-$ cross section extracted
for the $^4$He at the largest angle is very small.

Systematic uncertainties connected with the chosen cross section model
have be checked by using different modifications of the model
parameters and additionally by checking different FSI modifications of
the model.  For all targets the values obtained with the model are
very stable against small variations in the 3-6\%
range. Conservatively, we estimate model dependent uncertainties to
be within 6\%.

Figures~\ref{fig:3x4} and~\ref{fig:fit6_0} show some missing strength of 
the fit in the $\Lambda$ region for $A=3,4,12$. Integrating the data
as well as the fit in the low $M_X$ region below the $\Lambda$
threshold up to the $\Sigma^0$ threshold gives an estimate of the
relative missing strength. Nevertheless, we assume that our modeling
of the pure quasifree interaction is correct and that this additional
strength is due to FSI not described properly by our ERA - we thus 
assume that this additional strength will not modify the extracted
cross section for the quasifree production on these targets. We
estimated that at most 1/3 of the percentage of missing strength
tabulated in Table~\ref{tab:strength}
should be added to the systematic uncertainties of the cross section
values of Table~\ref{tab:values}.   
  
We checked the systematic uncertainty induced by the choice of a
particular YN interaction potential within the ERA applied. Again, we
see strong dependences on the angle for either target. As an example,
the quasifree $^4\mathrm{He}(e,e'K^+)\Lambda$ cross section changes by
5\% to 6\% if the Nijmegen 97f or the J\"ulich A hyperon nucleon
potential are used within the above mentioned effective range ansatz.
This change of the cross section then influences the extraction of the
quasifree $^4\mathrm{He}(e,e'K^+)\Sigma^0$ cross section by +2.7\% to
-2.6\% respectively. Values for $\Lambda$ and $\Sigma^0$ do not show a
strong angle dependence here, values for the $A=2,3$ targets are in
similar range. Introducing final state interaction for the $\Sigma^-$,
however, may change the cross section for $\Sigma¯$ by up to 100\%
compared to the value obtained without using final state interaction.
However the fits without final state interaction are of far lesser
quality than the ones including final state interaction. We, therefore,
do not consider them in Table~\ref{tab:values}.

\section*{Effective proton number}
Following Ref~\cite{Yamazaki:1995bc}, an effective proton
number $Z_{\mathrm{eff}}$ may be obtained by comparing the nuclear
with the elementary cross section for $\Lambda$ production:
\begin{equation}
  \label{eq:9}
  \left(\frac{d^2\sigma}{d\Omega}\right)_A =
  Z_{\mathrm{eff}} \left(\frac{d^2\sigma}{d\Omega}\right)_{\mathrm{H}}.
\end{equation}
In this ansatz we have to correct for final state interaction by
dividing the cross sections by the respective enhancement factors of
Table~\ref{tab:fsi}.  If we restrict ourselves to normalizing the
respective $\Lambda$ distribution for the nuclear targets by the
$\Lambda$ distribution from hydrogen, i.e.  $Z_{\mathrm{eff}}\simeq
\sigma_{\Lambda}(A)/\sigma_{\Lambda}(^1\mathrm{H})$, we obtain for the
near parallel kinematics and full $\phi$ coverage effective proton
numbers as given in Table~\ref{tab:zeff}. For helium, these numbers are in nice
agreement with phenomenological estimates of the respective effective
proton numbers that are derived with a procedure similar to that
presented in Ref.~\cite{Lee:1997mt}. The authors of \cite{Lee:1997mt} determine the effective
proton number in photoproduction of $\Lambda$ hyperons on carbon via
an eikonal approximation, where the thickness function $T$ is taken
to be a harmonic oscillator wave function. The integral (eq.~(22) of
Ref.~\cite{Lee:1997mt})
\begin{equation}\label{eq:z_eff}
Z_{\mathrm{eff}}= \frac{\pi}{2} \int dx T(x) \exp\left[
-\frac{\sigma^{\mathrm{tot}}_{\gamma
    N}+\sigma^{\mathrm{tot}}_{KN}}{2}T(x)\right]
\end{equation}
may then be calculated analytically, using $\sigma^{\mathrm{tot}}_{\gamma
    N}=0.2$ mb and $\sigma^{\mathrm{tot}}_{KN}=12$ mb.
Using only s-waves,
eq.~(\ref{eq:z_eff}) furthermore reduces to 
\begin{gather}\label{eq:21}
  Z_{\mathrm{eff}}= a\left(1-\exp\left(-\frac{Z}{a}\right)\right);\\
  a=\frac{\pi b^2}{\sigma},\;\sigma=\sigma^{\mathrm{tot}}_{\gamma
    N}+\sigma^{\mathrm{tot}}_{KN};X\nonumber\\
    T(x)=\frac{2Z}{\pi b^2}\exp\left(-\frac{x}{b^2}\right).
\end{gather}
For estimating the effective proton number for our targets, we follow
this approach: for $^4$He we take the rms charge radius of $^4$He from
literature and fit parameter $b=1.32$ fm\footnote{T.S.H.~Lee private
  communication.}. For $^3$He we extrapolate the fit parameter $b$
from the values from $^4$He. For carbon, the values of Ref.~\cite{Lee:1997mt} are used. Note that using eq.~(\ref{eq:21}), i.e.\
not taking into account p-wave contributions for carbon, would
yield an effective proton number of $~4.0$ instead of $~4.1$. 
Table~\ref{tab:zeff} summarizes our
estimates and experimentally derived values.  For the
deuteron we also estimated $Z_{\mathrm{eff}}$ by using a Hulth\'en
wave function for the deuteron\footnote{A.~Titov, private
  communication.},
\begin{gather}
\psi(r) = \frac{u(r)}{r};\quad u(r)=N\left(e^{-\alpha r}-e^{-\beta
    r}\right);\\
N= \sqrt{\frac{\alpha\beta(\alpha+\beta)}{2\pi(\alpha-\beta)^2}};\\
\alpha=0.2316\,\mathrm{fm}^{-1},\, \beta=1.268\,\mathrm{fm}^{-1}\nonumber
\end{gather}
for which we obtain $Z_{\mathrm{eff}}^{D}\simeq 0.88 $ by numerically
integrating (\ref{eq:z_eff}). 

The overall results are in fair agreement with the estimated
values. The value for carbon seems a bit high, but this 
probably reflects the rather poor statistics of carbon, and the
difficulty of modeling the cross section and FSI in heavier nuclei. 

\section*{Deeply bound kaonic states}
From kaon physics many indications were reported that the $\bar{K}N$
nuclear potential is
attractive~\cite{Batty:1997zp,Laue:1999yv,Scheinast:2004xy}.
Predictions of the depths of such potentials vary, as does the possibility
of producing deeply bound kaonic states in nuclei. Predictions
conclude that such a system should have a drastically contracted core
with simple core radius roughly 1/2 of the normal core size, i.e.\
without the bound $\bar{K}$.  It is suggested that a kaonic nuclear
system, e.g.\ $K^-ppn$ would decay into $\Lambda p n$ via the $K^-
pp(n) \to \Lambda p(n)$ and a $\Lambda^*(1405)$ doorway state. The
decay products should be visible in several
reactions~\cite{Yamazaki:2002uh}, among which also is
electroproduction on light nuclei.

Recently, several groups have searched for these states in light
nuclei.  Such states,
Refs~\cite{Friedman:1999rh,Friedman:1999pu,Akaishi:2002bg,Yamazaki:2002uh,Dote:2002db},
are predicted to imply potential depths of $\sim 100$ MeV and more
while showing small widths of $\sim$10--60~MeV. Some experimental
evidence was reported from ${}^4\mathrm{He}(\mathrm{stopped}\;K^-,p)$
experiments at KEK~\cite{Suzuki:2004ep,Iwasaki:2003kj}, from in-flight
$^{16}\mathrm{O}(K^-,n)$ experiments at AGS~\cite{Kishimoto:2005na} as
well as from the FINUDA experiment at DA$\Phi$NE~\cite{Agnello:2005qj}
in $pp\to \Lambda p$ invariant mass spectroscopy. For a criticism of
the interpretation of these data as bound kaonic states
see Ref.~\cite{Oset:2005sn}. Moreover, in a recent
publication~\cite{Shevchenko:2006xy}, a width estimate, obtained by
means of a Faddeev calculation for a $K^-pp$ quasi-bound state, is of
the order of 90-110 MeV, a result at variance with the results of the
FINUDA experiment~\cite{Agnello:2005qj}.

Experiment E91-016 may access inclusive distributions of final states
which may be decay channels of the presumed bound states
(cf.~\cite{Akaishi:2002bg}) for $A=2$: $p+\Lambda$, $n+\Lambda$;
$A=3$: $p+p+\Lambda$, $d+\Lambda$; $A=4$: $\Lambda + t$, 
$\Lambda + {}^3$He.  Taking the values of the presumed states
from Ref.~\cite{Yamazaki:2002uh} and comparing with
Figs~\ref{fig:3x4} and~\ref{fig:fit6_0}, we find that for $A=2$ we
are very much at the edge of the acceptance ($M_{ppK^-}\sim
2.32$~GeV), whereas for $A=3$ ($M_{pppK^-}\sim M_{ppnK^-}
\sim M_{pnnK^-}\sim 3.1$~GeV) the presumed states are well
within the acceptance, for $A=4$ we also should be within the
acceptance ($M_{pppnK^-} \sim M_{ppppK^-}\sim 4.13$~GeV).
However, while we do expect to have sensitivity within our acceptance
for the $A=3,4$ cases, the $M_X$ distributions for all nuclei are well
described by our model of quasifree kaon production from nucleons distributed
according to a theoretical spectral function.  Our experiment does not
show evidence for deeply bound kaonic states visible in
electroproduction, as was proposed in Ref.~\cite{Akaishi:2002bg}.

\section*{Summary}
This paper presented for the first time results on the cross section,
angular distributions, and nuclear dependence of kaon
electroproduction from hydrogen, deuterium, helium-3, helium-4, and
carbon. As a result we obtain quasifree distributions for the
respective $\Lambda$, $\Sigma^0$ and $\Sigma^-$ hyperons, which are
reconstructed by missing mass techniques. These quasifree angular
distributions show a behavior similar to the distributions obtained on
the free proton.  For the extraction of the respective cross sections the
dedicated simple model that was used gave the best description of the
data over the kinematic range of the experiment.  The extraction of
cross sections relied on three decisive steps:\ using a model developed
for the electroproduction of open strangeness on the free proton; employing
this model for the description of the quasifree process on
nuclei; and using spectral functions convolved with the elementary model.
Moreover, it is mandatory to include final state interaction in the
vicinity of the respective thresholds for the production of $\Lambda,
\Sigma^0$, and $\Sigma^-$.  Final state interactions are modeled by an
effective range approximation using hyperon nucleon potentials.  For
carbon, we clearly see the $^{12}_{\Lambda}$B bound state, which we do
not resolve further, but for which we give a cross section estimate.

Effective proton numbers are extracted by comparing the nuclear cross
section with the cross section on the free proton. Correcting for
final state interaction we see the measured nuclear effects for $A=2,
3, 4$ in accordance with estimates using a simple eikonal
approximation. For carbon, our numbers are higher than the estimated
effective proton numbers, which might be due to the small data set at
hand.

The missing mass distribution for helium do not show any noticeable
structures in the vicinity of $M_X \sim 3.1 $ GeV for $^3$He
or $M_X \sim 4.13$ GeV for $^4$He such that no supportive
evidence for deeply bound kaonic states may be drawn from these
distributions. It should be pointed out again that these measurements
are inclusive and that an exclusive measurement may still have more
power in making a statement on these postulated bound states. 

Electroproduction experiments with high intensity beams on light nuclear  
targets are a fascinating subject which will be studied further at
Jefferson Laboratory~\cite{Nakamura:2003us} and MAMI-C at
Mainz~\cite{Pochodzalla:2004dv}.  

\begin{acknowledgments}
  We would like to thank Dr.\ T.S.H.\ Lee, Dr.\ A.\ Titov and Dr.\
  H.-W.\ Barz for helpful discussions and support on calculating the
  effective proton numbers.  This work was supported in part by the
  U.S. DOE under contract Nos.\ DE-AC03-06CH11357 (ANL),
  DE-AC05-84ER40150 (JLab), and the NSF\@.  The excellent support of
  the staff of the Accelerator and Physics Division of JLab is
  gratefully acknowledged.  F.D.\ acknowledges the support by the
  A.v.Humboldt-Stiftung and the support by ANL for hosting this
  research.
\end{acknowledgments}



\cleardoublepage
\begin{table}[htb]
  \centering
  \begin{tabular}{l|ccc}
      & $c_1$ & $c_2$ & $c_3$ \\ \hline
    $f(Q^2)$ &  $0.430\;\mathrm{\mu b/sr}$  &  &  \\
    $g(W)$ &  $4.470\;\mathrm{MeV}^2$ & $0.00089\;\mathrm{MeV}^2$ & $0.0062\;\mathrm{MeV^2}$ \\
    $h(\Delta t)$ & $-2.14$ &   &  \\
    $i(\phi)$ &  $0.438$ & $-0.048$ & $0.008$ \\
  \end{tabular}
  \caption{Fit parameters for the model cross section for
    ${}^1\mathrm{H}(e,e'K^+)\Lambda$ from \cite{Cha:2000xy}.}
  \label{tab:jinparams}
\end{table}

\cleardoublepage

 \begin{table}[hbtp]
  \begin{tabular}{lp{1.2cm}p{1cm}p{1cm}p{1cm}}
    \raisebox{-2.5ex}[0cm][0cm]{target} & \raisebox{-2.5ex}[0cm][0cm]{angle$({^\circ})$} & \multicolumn{3}{c}{$Y_{\mathrm{FSI}}/Y_{\mathrm{no\_FSI}}$}    \\
            &                &  $\Lambda$ & $\Sigma^0$ & $\Sigma^-$ \\\hline

    ${}^2\mathrm{H}$     &  $1.7$  & 4\%   & 3\%  & 2\%  \\ 
    ${}^3\mathrm{He}$    &  $1.7$  & 15\%  & 8\%  & 10\%  \\ 
    ${}^3\mathrm{He}$    &  $6$    & 12\%  & 7\%  & 9\%  \\ 
    ${}^3\mathrm{He}$    &  $12$   & 9\%   & 5\%  & 7\%  \\ 
    ${}^4\mathrm{He}$    &  $1.7$  & 13\%  & 7\%  & 11\%  \\ 
    ${}^4\mathrm{He}$    &  $6$    & 12\%  & 6\%  & 9\%  \\ 
    ${}^4\mathrm{He}$    &  $12$   & 7\%   & 4\%  & 10\%  \\ 
    ${}^{12}\mathrm{C}$  &  $1.7$  & 10\%  & 5\%  & 8\%  
  \end{tabular}
  \caption{Final State interaction enhancement factors. The factor is
    ratio of the integrated yield of missing mass spectra before $Y_{\mathrm{no\_FSI}}$ and
    after $Y_{\mathrm{FSI}}$ applying the final
    state contribution for the respective kinematic setting and
    target. The integration is carried out over the kinematic range for
    the respective targets, cf.\ Figs.~\ref{fig:3x4} and~\ref{fig:fit6_0}.}
  \label{tab:fsi}
\end{table}

\cleardoublepage

\begin{sidewaystable}
{\scriptsize
  \caption{\label{tab:values} Differential cross sections for
    electroproduction of $K^+\Lambda$, $K^+\Sigma^{0,-}$ final states
    on $A=1,2,3,4,12$ targets. A prescription for separating the
    $\Sigma^0$, $\Sigma^-$ cross sections is discussed in the text.
    Independently a combined $K^+\Sigma$ cross section is given.
    $\sigma_{\mathrm{lab}}$ denotes the five fold laboratory
    differential cross section ${d^5\sigma}/{d\Omega_e dE_e d\Omega_K
    }$ (in \ensuremath{\mathrm{( nb/GeV/sr^2)}}).  $\sigma_{\mathrm{cm}}$ denotes the two fold differential
    cross section ${d^2\sigma}/{d\Omega}$ (in \ensuremath{\mathrm{({\mu}b/sr)}})in the virtual
    photon--nucleus center of mass system. Uncertainties given include
    the combined statistical and fitting uncertainties. Uncertainties
    from Table~\ref{tab:certain} have to be added to these values.
    The first row shows data for $1.7{^\circ}$ averaged over the azimuth.
    A $9\%$ systematic error has to be added to the cross sections
    given, see text and Table~\ref{tab:certain}. Note that values are 
    not given per contributing nucleon, cf.\ text.
  }
\begin{ruledtabular}
  \begin{tabular}{ccccccccccc}
\multicolumn{1}{c}{Target} & 
\multicolumn{2}{c}{\ensuremath{^1\mathrm{H }}} & 
\multicolumn{2}{c}{\ensuremath{^2\mathrm{H }}} & 
\multicolumn{2}{c}{\ensuremath{^3\mathrm{He}}} & 
\multicolumn{2}{c}{\ensuremath{^4\mathrm{He}}} & 
\multicolumn{2}{c}{\ensuremath{^{12}\mathrm{C}}} 
\\\hline
\multicolumn{1}{c}{\ensuremath{\theta_{\gamma^*, K^+}^{\mathrm{lab}}(^{\circ})}} & 
\multicolumn{1}{c}{\ensuremath{\sigma_{\mathrm{lab}}}}  & 
\multicolumn{1}{c}{\ensuremath{\sigma_{\mathrm{cm}}}}   & 
\multicolumn{1}{c}{\ensuremath{\sigma_{\mathrm{lab}}}}  & 
\multicolumn{1}{c}{\ensuremath{\sigma_{\mathrm{cm}}}}   & 
\multicolumn{1}{c}{\ensuremath{\sigma_{\mathrm{lab}}}}  & 
\multicolumn{1}{c}{\ensuremath{\sigma_{\mathrm{cm}}}}   & 
\multicolumn{1}{c}{\ensuremath{\sigma_{\mathrm{lab}}}}  & 
\multicolumn{1}{c}{\ensuremath{\sigma_{\mathrm{cm}}}}   & 
\multicolumn{1}{c}{\ensuremath{\sigma_{\mathrm{lab}}}}  & 
\multicolumn{1}{c}{\ensuremath{\sigma_{\mathrm{cm}}}}   
\\
\multicolumn{11}{c}{\ensuremath{\Lambda}}\\
\ensuremath{\langle1.7\rangle} &
10.6 \ensuremath{\pm} 0.2 	&	 0.47 \ensuremath{\pm} 0.01 	&	 
9.4 \ensuremath{\pm}  0.2 	&	 0.41 \ensuremath{\pm} 0.01 	&	 
21.7 \ensuremath{\pm} 0.2 	&        0.95 \ensuremath{\pm} 0.01 	&
19.8 \ensuremath{\pm} 0.2 	&	 0.86 \ensuremath{\pm} 0.01        &
61.6 \ensuremath{\pm} 1.5  &        2.64 \ensuremath{\pm} 0.07        
\\
\ensuremath{1.7}&
9.8 \ensuremath{\pm} 0.4 	&	 0.43 \ensuremath{\pm} 0.03 	&
9.4 \ensuremath{\pm} 0.5 	&	 0.41 \ensuremath{\pm} 0.02 	&
20.4 \ensuremath{\pm} 0.3 	&        0.89 \ensuremath{\pm} 0.02 	&
18.2 \ensuremath{\pm} 0.3 	&	 0.79 \ensuremath{\pm} 0.02        &
                &                               
\\
\ensuremath{6}  &
9.9 \ensuremath{\pm} 0.1 	&	 0.44 \ensuremath{\pm} 0.01 	&
                &          	              	&
19.5 \ensuremath{\pm} 0.3 	&	 0.87 \ensuremath{\pm} 0.02 	&
17.7 \ensuremath{\pm} 0.3 	&	 0.79 \ensuremath{\pm} 0.03        &
                &                               
\\
\ensuremath{12} &
7.6 \ensuremath{\pm} 0.1 	&	 0.36 \ensuremath{\pm} 0.01 	&
              	&	                        &
15.0 \ensuremath{\pm} 0.5 	&        0.71 \ensuremath{\pm} 0.04        &
14.2 \ensuremath{\pm} 0.4 	&        0.67 \ensuremath{\pm} 0.02        &
                &                               
\\\hline
\multicolumn{11}{c}{\ensuremath{\Sigma^0}}\\
\ensuremath{\langle1.7\rangle} &
3.0 \ensuremath{\pm} 0.2 	&	 0.12 \ensuremath{\pm} 0.01 	&
2.8 \ensuremath{\pm} 0.2   &	 0.11 \ensuremath{\pm} 0.02 	&
6.4 \ensuremath{\pm} 0.2   &	 0.25 \ensuremath{\pm} 0.01 	&
6.3 \ensuremath{\pm} 0.1  	&	 0.25 \ensuremath{\pm} 0.01        &
20.7 \ensuremath{\pm} 0.5  &        0.81 \ensuremath{\pm} 0.02        
\\
\ensuremath{1.7}&
3.0 \ensuremath{\pm} 0.4 	&	 0.12 \ensuremath{\pm} 0.01 	&
3.0 \ensuremath{\pm} 0.2 	&	 0.12 \ensuremath{\pm} 0.01 	&
6.5 \ensuremath{\pm} 0.2 	&        0.26 \ensuremath{\pm} 0.01 	&
6.2 \ensuremath{\pm} 0.2 	&	 0.25 \ensuremath{\pm} 0.01        &
                &                               
\\
\ensuremath{6}  &
3.3 \ensuremath{\pm} 0.1 	&	 0.14 \ensuremath{\pm} 0.01 	&
              	&	                        &
6.8 \ensuremath{\pm} 0.2   &        0.27 \ensuremath{\pm} 0.01 	&
6.6 \ensuremath{\pm} 0.2 	&        0.24 \ensuremath{\pm} 0.01        &
                &                               
\\
\ensuremath{12} &
3.2 \ensuremath{\pm} 0.1 	&	 0.14 \ensuremath{\pm} 0.01 	&
                &	              	        &
6.6 \ensuremath{\pm} 0.2 	&	 0.29 \ensuremath{\pm} 0.01 	&
6.6 \ensuremath{\pm} 0.2 	&	 0.29 \ensuremath{\pm} 0.01        &
                &                               
\\\hline
\multicolumn{11}{c}{\ensuremath{\Sigma^-}}\\
\ensuremath{\langle1.7\rangle} &
               	&	              	        &
1.9 \ensuremath{\pm} 0.2   &	 0.08 \ensuremath{\pm} 0.01 	&
4.6 \ensuremath{\pm} 0.2  	&	 0.18 \ensuremath{\pm} 0.01        &
4.8 \ensuremath{\pm} 0.2  	&	 0.19 \ensuremath{\pm} 0.01        &
16.5 \ensuremath{\pm} 2.6  &        0.64 \ensuremath{\pm} 0.1         
\\
\ensuremath{1.7}&
               	&        	              	&
1.8 \ensuremath{\pm} 0.5   &	 0.07 \ensuremath{\pm} 0.02 	&
3.9 \ensuremath{\pm} 0.4   &	 0.15 \ensuremath{\pm} 0.02 	&
4.6 \ensuremath{\pm} 0.4   &	 0.18 \ensuremath{\pm} 0.02        &
                &                               
\\
\ensuremath{6}  &
              	&        	              	&	                     
        	&	              	        &
3.5 \ensuremath{\pm} 0.5  	&	 0.14 \ensuremath{\pm} 0.02 	&
2.3 \ensuremath{\pm} 0.6   &	 0.09 \ensuremath{\pm} 0.02        &
                &                               
\\
\ensuremath{12} &
               	&	                 	&	                 
          	&	                	&
3.3 \ensuremath{\pm} 1.2  	&	 0.14 \ensuremath{\pm} 0.05 	&
0.3 \ensuremath{\pm} 0.6  	&	 0.01 \ensuremath{\pm} 0.02        &
                &                               
\\\hline
\multicolumn{11}{c}{\ensuremath{\Sigma}}\\
\ensuremath{\langle1.7\rangle} &
                    	&	               	&
4.7 \ensuremath{\pm} 0.2 	&	 0.19 \ensuremath{\pm} 0.02 	&
10.5 \ensuremath{\pm} 0.2 	&        0.42 \ensuremath{\pm} 0.02 	&
11.1 \ensuremath{\pm} 0.3 	&	 0.44 \ensuremath{\pm} 0.02        &
37.0 \ensuremath{\pm} 2.6  &        1.45 \ensuremath{\pm} 0.10        
\\
\ensuremath{1.7} &
               	&	                 	&
4.9 \ensuremath{\pm} 0.6 	&	 0.19 \ensuremath{\pm} 0.02 	&
9.9 \ensuremath{\pm} 0.4 	&        0.39 \ensuremath{\pm} 0.02 	&
10.8 \ensuremath{\pm} 0.5 	&	 0.43 \ensuremath{\pm} 0.02        &
                &                               
\\
\ensuremath{6} &
                &        	               	&
                &	                 	&
9.7 \ensuremath{\pm} 0.4 	&	 0.39 \ensuremath{\pm} 0.02 	&
9.0 \ensuremath{\pm} 0.6 	&	 0.36 \ensuremath{\pm} 0.02        &
                &                               
\\
\ensuremath{12} &
               	&       	              	&
                &	                 	&
9.3 \ensuremath{\pm} 0.9 	&	 0.40 \ensuremath{\pm} 0.04 	&
7.0 \ensuremath{\pm} 0.2 	&	 0.30 \ensuremath{\pm} 0.02        &
                &                               
\\
\end{tabular}
\end{ruledtabular}
}
\end{sidewaystable}

\cleardoublepage
\begin{table}[hbtp]
  \centering
  \begin{tabular}{ccccc}
    target & $\langle1.7^{{^\circ}}\rangle$& $1.7^{{^\circ}}$ & $6^{{^\circ}}$ &
    $12^{{^\circ}}$ \\
    $^2$H  & 0.3\%   & 0.3\%        &           &     \\
    $^3$He & 0.7\%   & 2.3\%        &   3\%     & 8\% \\
    $^4$He & 5  \%   & 6  \%        &   7\%     & 18\%\\
    $^{12}$C & 22\%   & 
  \end{tabular}
  \caption{Missing relative strength in low-mass $\Lambda$ region, integrated
    up to the lowest lying $\Sigma^0$ threshold. These values were
    obtained for the choice of our cross section model
    (\ref{eq:2}-\ref{eq:6}) and Nijmegen YN potential as discussed in the text.}
  \label{tab:strength}
\end{table}

\cleardoublepage

\begin{table}[hbtp]
  \centering
  \begin{tabular}{cc}
    type & uncertainty ($\%$) \\\hline
    experimental systematics & 6\% \\
    cross section model & 6\% \\
    FSI model & 3\% \\\hline
    total & 9\% 
  \end{tabular}
  \caption{Breakdown of systematic uncertainties. These uncertainties have to be added to the uncertainties
given in Table~\ref{tab:values}.}
  \label{tab:certain}
\end{table}

\cleardoublepage

\begin{table}[htbp]
  \centering
  \begin{tabular}{clcccll}
    target & \multicolumn{2}{c}{rms (fm)} & b (fm) & $Z$ & \multicolumn{1}{c}{$Z_{\mathrm{eff}}^{\mathrm{exp}}$} & \multicolumn{1}{c}{$Z_{\mathrm{eff}}^{\mathrm{est}}$}\\\hline
    $^2$H & 2.140& \cite{NIST:2007} & (1.71$^{*}$) & 1 &0.85 $\pm$ 0.09 &
    0.89 (0.93$^{*}$) \\
    $^3$He     & 1.976&\cite{Ottermann:1985km,TUNL:2007}  & 1.58 & 2 &
    1.76 $\pm$ 0.16 & 1.7\\
    $^4$He     & 1.647&\cite{Ottermann:1985km,TUNL:2007}  & 1.32 & 2 &
    1.61 $\pm$ 0.16 & 1.6\\
    $^{12}$C  & 2.483&\cite{Lee:1997mt} & 1.64   & 6 & 5.15 $\pm$ 0.7 &
    4.1 
  \end{tabular}
  \caption{Effective proton numbers derived from the cross section in
    Table~\ref{tab:values} and estimates of effective proton numbers,
    derived from the calculated absorption taking rms charge radii from literature and other
    references, cf.\ text. The superscript
    $^{*}$ denotes a harmonic oscillator function. Values are given
    for data at 1.7$^{{^\circ}}$, averaged over the azimuth.}
  \label{tab:zeff}
\end{table}


\cleardoublepage

\cleardoublepage

\cleardoublepage

\cleardoublepage

\cleardoublepage

\cleardoublepage

\cleardoublepage

\end{document}